\documentclass[
amssymb,nofootinbib,superscriptaddress,longbibliography,notitlepage,twocolumn
]{revtex4-2}

\usepackage{physics}
\usepackage{graphicx}
\usepackage{dcolumn}
\usepackage{bm}
\usepackage{hyperref}
\usepackage{algorithm}
\usepackage{algpseudocode}
\usepackage{amsmath,amsfonts,amssymb,cases}
\usepackage{bm}
\usepackage{qcircuit}
\usepackage{mathtools}
\usepackage{bbm}
\usepackage{empheq}

\begin{document}

\title{Pricing multi-asset derivatives by variational quantum algorithms}

\author{Kenji Kubo}
\email{kenjikun@mercari.com}
\affiliation{R4D, Mercari Inc., Roppongi Hills Mori Tower 18F, 6-10-1, Roppongi, Minato-ku, Tokyo 106-6118, Japan}
\affiliation{Graduate School of Engineering Science, Osaka University, 1-3 Machikaneyama, Toyonaka, Osaka 560-8531, Japan.}

\author{Koichi Miyamoto}%
 \email{miyamoto.kouichi.qiqb@osaka-u.ac.jp}
 \affiliation{Center for Quantum Information and Quantum Biology, Osaka University, 1-2 Machikaneyama, Toyonaka, Osaka 560-0043, Japan.}

\author{Kosuke Mitarai}
\email{mitarai@qc.ee.es.osaka-u.ac.jp}
\affiliation{Graduate School of Engineering Science, Osaka University, 1-3 Machikaneyama, Toyonaka, Osaka 560-8531, Japan.}
\affiliation{Center for Quantum Information and Quantum Biology, Osaka University, Japan.}
\affiliation{JST, PRESTO, 4-1-8 Honcho, Kawaguchi, Saitama 332-0012, Japan.}

\author{Keisuke Fujii}
\email{fujii@qc.ee.es.osaka-u.ac.jp}
\affiliation{Graduate School of Engineering Science, Osaka University, 1-3 Machikaneyama, Toyonaka, Osaka 560-8531, Japan.}
\affiliation{Center for Quantum Information and Quantum Biology, Osaka University, Japan.}
\affiliation{RIKEN Center for Quantum Computing, Wako Saitama 351-0198, Japan}

\date{\today}

\begin{abstract}
    Pricing a multi-asset derivative is an important problem in financial engineering, both theoretically and practically. 
    Although it is suitable to numerically solve partial differential equations to calculate the prices of certain types of derivatives, the computational complexity increases exponentially as the number of underlying assets increases in some classical methods, such as the finite difference method.
    Therefore, there are efforts to reduce the computational complexity by using quantum computation. 
    However, when solving with naive quantum algorithms, the target derivative price is embedded in the amplitude of one basis of the quantum state, and so an exponential complexity is required to obtain the solution. 
    To avoid the bottleneck, the previous study~[Miyamoto and Kubo, IEEE Transactions on Quantum Engineering, \textbf{3}, 1--25 (2022)] utilizes the fact that the present price of a derivative can be obtained by its discounted expected value at any future point in time and shows that the quantum algorithm can reduce the complexity.
    In this paper, to make the algorithm feasible to run on a small quantum computer, we use variational quantum simulation to solve the Black-Scholes equation and compute the derivative price from the inner product between the solution and a probability distribution.
    This avoids the measurement bottleneck of the naive approach and would provide quantum speedup even in noisy quantum computers.
    We also conduct numerical experiments to validate our method. 
    Our method will be an important breakthrough in derivative pricing using small-scale quantum computers.
\end{abstract}
\maketitle

\section{Introduction}\label{sec: introduction}
Quantum computers actively utilize quantum phenomena to solve large-scale problems that could not be performed with conventional classical computers.
In recent years, applications of quantum computers have been discussed in financial engineering.
Specifically, the applications include portfolio optimization~\cite{hodson2019portfolio,rebentrost2018quatnum,kerenidis2019quantum}, risk measurement~\cite{miyamoto2020reduction,woerner219quantum,egger2021credit,kaneko2021quantum,miyamoto2022quantumrisk}, and derivative pricing~\cite{rebentrost2018quantum,martin2021toward,stamatopoulos2020option,ramoscalderer2021quantum,fontanela2021quantum,radha2021quantum,gonzalezconde2021pricing,carrera2021efficient,tang2021quantum,chakrabarti2021thresholdquantum,an2021quantumaccelerated,kaneko2022quantum,miyamoto2022bermudan,alghassi2022variational}.
Comprehensive reviews of these topics are presented in Refs. \cite{egger2020quantum,orus2019quantum,bouland2020prospects,herman2022quantum}.
    
Among these applications, we consider the pricing of derivatives.
Derivatives are the products that refer to the prices of underlying assets such as stocks, bonds, currencies, etc., and their payoff depends on the prices of the assets.
For example, a European call option, one of the simplest derivatives, has a predetermined maturity $T>0$ and strike price $K$, and its holder gets paid back $\max(S(T)-K, 0)$ for the asset price $S(T)$ at $T$.
For such a simple derivative, the theoretical price can be computed analytically in some models such as the Black-Scholes (BS) model \cite{black1973pricing}.
If one wishes to calculate prices for derivatives with more complex payoffs, numerical calculations are required~\cite{hull2012options}.

There are many algorithms for numerical calculations. For the pricing of certain types of derivatives, such as barrier options, it is suitable to solve the partial differential equations (PDE) called Black-Scholes PDE (BSPDE)~\cite{shreve2004stochastic} by discretizing them using the finite difference method (FDM).
However, in the case of multi-asset derivatives, the number of grid points increases exponentially with respect to the number of referenced assets, making price calculation difficult.
When the number of assets is $d$ and the number of grid points is $n_{\mathrm{gr}}$ for one asset, the total number of grid points is $n_{\mathrm{gr}}^d$.
If we take $n_{\mathrm{gr}}$ in proportion to $\epsilon^{-1/2}$ to achieve the error level $\epsilon$ (see Lemma II.1 in \cite{miyamoto2022pricing}), classical FDM requires the computational complexity of $O((1/\epsilon)^{O(d)})$.

To overcome this difficulty, several methods~\cite{fontanela2021quantum,gonzalezconde2021pricing,radha2021quantum,alghassi2022variational} have been proposed to efficiently solve the BSPDE using quantum computers.
However, when solving the discretized BSPDE with these quantum algorithms, the target derivative price is embedded in the amplitude of one basis of the resulting quantum state, so it requires exponentially large computational complexity to extract it as classical information.
Ref. \cite{miyamoto2022pricing} has shown that the complexity can be substantially reduced using the fact that the present derivative price can be calculated as the expected value of the discounted derivative price at a future point in time.
They calculate the inner product of the state in which the future derivative prices are embedded and the state in which the probability distribution is embedded using the quantum amplitude estimation (QAE)~\cite{brassard2002quantum}.
Instead of retrieving one of the amplitudes of the output state of the quantum algorithm, the present price of the derivative can be efficiently calculated since all of the amplitudes can be used.
In fact, the complexity of the method proposed in Ref.~\cite{miyamoto2022pricing} does not have a factor like $(1/\epsilon)^{O(d)}$, but has only $\mathrm{poly}(1/\epsilon, d)$.
This means that their method has substantial speedup compared to the classical FDM.

However, it should be noted that their method is constructed on the quantum ordinary differential equation (ODE) solver~\cite{berry2017quantum} and the QAE, which requires a large-scale quantum computer with error correction.
In addition, it is assumed that we are given the oracle that generates a quantum state in which the boundary conditions of the BSPDE are encoded in amplitudes.
As the derivatives are currently dealt with in practice, it is desirable to calculate derivative prices even with a small-scale quantum computer closer to realization.

In this paper, we propose a variational quantum algorithm for pricing multi-asset derivatives.
This is the way to exploit the essential feature proposed in Ref.~\cite{miyamoto2022pricing} with variational quantum algorithms and hence thought to work with near-term quantum computers.
Our algorithm has the following three parts; embedding the probability distribution of the underlying asset prices into the quantum state, solving the BSPDE with boundary conditions, and calculating the inner product.
For the first part, we can use the quantum generative algorithms~\cite{zoufal2019quantum,situ2020quantum,lloyd2018quantum, dallaire2018quantum,kyriienko2022protocols} or variational quantum simulation (VQS) for the Fokker-Planck equations~\cite{endo2020variational,yuan2019theoryofvariational,cerezo2021variational,kubo2021variational,alghassi2022variational}, which describe the time evolution of the probability density functions of the stochastic processes.
For the second part, we discretize the BSPDE using the FDM and solve it using VQS.
For the third part, we evaluate the square of the inner product of the states, obtained by the first and the second parts of our method, using the SWAP test~\cite{carlos2013swap}.
Taking the square root of the output of the SWAP test and discounting by the interest rate, we obtain the present price of the derivative.
Although there is no guarantee of overall computational complexity due to the heuristic nature of the variational algorithm, we show that the number of measurements of the SWAP test has no factors like $(1/\epsilon)^{O(d)}$, which means that our method can avoid the bottleneck of retrieving derivative prices from the quantum state.
Since our algorithm requires quantum circuits with $O(\mathrm{poly}(d\log(1/\epsilon))$ few-qubit gates, even a small-scale quantum computer would be able to perform derivatives pricing with our method.
We perform numerical calculations for a single asset double barrier option and confirm that our method is feasible.

This paper is organized as follows.
Sec.~\ref{sec:Preliminary} is the preliminary section.
The notations in this paper are listed in Sec.~\ref{subsec: notation}.
We summarize the related works in Sec.~\ref{subsec: related work}.
In Sec.~\ref{subsec:derivative pricing} we introduce derivative pricing using the BSPDE with boundary conditions.
We also introduce FDM to discretize the BSPDE and obtain an ODE in Sec.~\ref{subsec:FDM}.
Sec.~\ref{subsection: VQS} gives an introduction to VQS, which is an algorithm for solving the ODE.
Sec.~\ref{sec: approximation} introduces the fact that the present price of the derivative can be approximated by the expected value of the future price.
In Sec.~\ref{sec:method}, we describe the proposed method.
We estimate the number of measurements required by the SWAP test in Sec.~\ref{subsec: measurement on SWAP test} and the whole time complexity of the proposed method in Sec.~\ref{subsec: complexity}.
We show the feasibility of our method through numerical simulations in Sec.~\ref{sec: numerical results}.
Conclusions are given in Sec. ~\ref{sec: conclusion}.

\section{Preliminary}\label{sec:Preliminary}
\subsection{Notation}\label{subsec: notation}
Here, we introduce the notation used in this paper.
We define $\mathbb{R}_+$ as a set of all positive real numbers, and for a positive integer $d$, $\mathbb{R}_+^d$ as a $d$-times direct product of $\mathbb{R}_+$.
For a positive integer $n$, $[n]\coloneqq\{1,2,\dots,n\}$.
For $\bm{v}=(v_1,v_2,\dots,v_n)^\top\in\mathbb{R}^n$, where $n$ is an integer not less than 2, and $i\in[n]$, we define $\bm{v}_{\wedge i}\in\mathbb{R}^{n-1}$ as a vector which is made by removing an element $v_i$ from $\bm{v}$, that is, $\bm{v}_{\wedge i} \coloneqq (v_1, v_2, \dots, v_{i-1}, v_{i+1}, \dots, v_{n})$.
We define the Euclidean norm for a vector $\bm{v}$ as $\|\bm{v}\|=\sqrt{\sum_i v_i^2}$.
For an integer $i$, we define $\ket{i}$ as one of the computational basis states with a binary representation of $i$ and for a vector $\bm{y}=(y_1,y_2,\dots,y_n)^\top\in\mathbb{C}^d$, we denote $\ket{\bm{y}}$ as an unnormalized state where the elements of $\bm{y}$ are encoded in the amplitudes, that is, $\ket{\bm{y}}\coloneqq\sum_{i=1}^{n}y_i\ket{i}$.


\subsection{Related work}\label{subsec: related work}
In this subsection, we explain the existing algorithms for solving the BSPDE with quantum computers.
Ref.~\cite{gonzalezconde2021pricing} transforms the BSPDE into a Schr\"{o}dinger equation, discretize the Hamiltonian by FDM, and solves it by diagonalization of discretized momentum operator with a quantum Fourier transformation. Refs.~\cite{radha2021quantum,fontanela2021quantum,alghassi2022variational} solve the discretized Schr\"{o}dinger equation by VQS, which is a variational quantum algorithm for solving ODEs. 
In the previous studies mentioned above,
the time complexity required to solve the BSPDE depends on the grid points only logarithmically.
However, there is still a problem that cannot be overlooked; extracting the calculated result from quantum computers may take exponentially long time with respect to $d$.
Solving the BSPDE from the maturity $(t=T)$ to the present $(t=0)$ with these quantum algorithms yields unnormalized state $\ket{\bm{V}(0)}$ whose elements are the derivative prices on the grid points  of underlying asset prices.
Note that, typically, we are interested in only one element of $\ket{\bm{V}(0)}$, the derivative price on the grid point corresponding to the present underlying asset prices.
However, since $\ket{\bm{V}(0)}$ has $O((1/\epsilon)^{O(d)})$ elements, the amplitude corresponding to $V_0$ in (normalized) $\ket{\bm{V}(0)}$ is exponentially small.
Therefore, the exponential time complexity is required to retrieve $V_0$ as classical information, and the quantum speedup will be lost.

Ref.~\cite{miyamoto2022pricing} shows the algorithm to overcome the problem.
They prepare the state $\ket{\bm{p}(t_\mathrm{ter})}$ in which the probability distribution of underlying asset prices on the grid points at a certain time $t_\mathrm{ter}\in[0,T]$ is embedded in the amplitudes.
Then, they discretize the BSPDE using FDM, solve it not to $t=0$ but $t=t_\mathrm{ter}$ with quantum ODE solver, and obtain the state $\ket{\bm{V}(t_\mathrm{ter})}$.
The inner product of these quantum states, which can be obtained by QAE, corresponds to the expected value of the derivative price at $t_\mathrm{ter}$ by $E\left[V(t_\mathrm{ter})\right]\simeq\sum_{i\in\mathcal{G}}p_i(t_{\mathrm{ter}})V_i(t_{\mathrm{ter}})=\braket{\bm{p}(t_\mathrm{ter})}{\bm{V}(t_\mathrm{ter})}$.
Discounting this expected value by the risk-free interest rate yields the present price of the derivative~\cite{hull2012options}.

Our algorithm is a variational version of Ref.~\cite{miyamoto2022pricing}.
Instead of using the quantum ODE solver and QAE, we use VQS and the SWAP test, respectively.
This enables derivatives pricing by BSPDE to be realized on a small-sized quantum computer.


\subsection{Derivative pricing}\label{subsec:derivative pricing}
To evaluate the price of a derivative, we need to model the dynamics of the prices of the underlying asset.
We adopt the BS model \cite{black1973pricing}, in which the prices of the underlying assets are assumed to follow geometric Brownian motions.
That is, we suppose that the prices of $d$ underlying assets at $t\in[0,T]$ are stochastic processes $\bm{S}(t)=(S_1(t), S_2(t), \dots, S_d(t))^\top\in\mathbb{R}_+^d$ that, under the risk-neutral measure, obey stochastic differential equations
\begin{align}\label{eq: geometric BM}
    dS_i(t) = rS_i(t)dt + \sigma_iS_i(t)dW_i(t).
\end{align}
Here, $r>0$ is the risk-free interest rate, $\sigma_i>0$ are volatility of the underlying assets, and they satisfy $0 < r < \frac{\sigma^2_i}{2}$ for all $i \in [d]$.
$W_i(t)$ are Brownian motions that satisfy $dW_idW_j=\rho_{i,j}dt, (i,j)\in[d]\times[d]$ with the correlation matrix $(\rho_{ij})_{1\le i,j \le d}$, which satisfies $\rho_{i,i}=1$ and $-1 < \rho_{i,j}=\rho_{j,i} < 1$ for $i\ne j$.

Derivatives are characterized by the payoff function $f_{\mathrm{pay}}$ at the maturity and the payoff conditions, which must be satisfied in order for the payoff to arise.
We describe the typical cases of the payoff functions and the payoff conditions later.
The price of the derivative is obtained as the conditional expected value of the payoff, conditioned on the price of the underlying assets, discounted by the risk-free rate~\cite{shreve2004stochastic}.
That is, given the underlying asset prices at time $t$ as $\bm{s}=(s_1, \dots, s_d)^\top\in\mathbb{R}_+^d$, and the payoff function at maturity $T$ as $f_{\mathrm{pay}}(\bm{S}(T))$, the price of the derivative is 
\begin{align}\label{eq: derivative pricing}
    V(t, \bm{s}) = E_Q\left[e^{-r(T-t)}f_{\mathrm{pay}}(\bm{S}(T))\mathbbm{1}_{\mathrm{NB}}\middle| \bm{S}(t)=\bm{s}\right],
\end{align}
where $E_Q$ is the expected value under the so-called risk-neutral measure.
Note that $\bm{S}(T)$ is a vector of random variable resulting from the time evolution of Eq.~\eqref{eq: geometric BM} from $t$ to $T$ with the condition $\bm{S}(t)=\bm{s}$.
$\mathbbm{1}_{\mathrm{NB}}$ is a random variable that takes $1$ if the payoff conditions are satisfied or $0$ otherwise.

The goal of derivative pricing is to find the present price of the derivative, that is, $V(0, \bm{s}_0)$, where $\bm{s}_0=(s_{1,0},\dots,s_{d,0})^\top\in\mathbb{R}_+^d$ is the present price of the underlying assets.
To this end, we use the BSPDE, which describes the time evolution of $V(t, \bm{s})$~\cite{shreve2004stochastic}.
That is, the derivative price $V(t, \bm{s})$ is the solution of the BSPDE
\begin{align}\label{eq:bspde}
    &\frac{\partial}{\partial t}V(t,\bm{s}) + \frac{1}{2} \sum_{i,j=1}^d \sigma_i \sigma_j s_i s_j\rho_{ij} \frac{\partial^2}{\partial s_i \partial s_j} V(t, \bm{s}) \nonumber \\ 
    &+ r\left( \sum_{i=1}^d s_i \frac{\partial}{\partial s_i} V(t, \bm{s})-V(t, \bm{s})\right)=0
\end{align}
on $[0,T)\times D$ with the boundary conditions
\begin{align}
    &V(T, \bm{s}) = f_{\mathrm{pay}}(\bm{s}), \\
    &V(t, (s_1,\dots,s_{i-1}, u_i, s_{i+1},\dots,s_d)^\top) \nonumber\\
    &\eqqcolon V_i^{\mathrm{UB}}(t, \bm{s}_{\wedge i}),~\mathrm{for}~i\in [d], \\
    &V(t, (s_1,\dots,s_{i-1}, l_i, s_{i+1},\dots,s_d)^\top) \nonumber\\
    &\eqqcolon V_i^{\mathrm{LB}}(t, \bm{s}_{\wedge i}),~\mathrm{for}~i\in [d].
\end{align}
where $u_i, l_i$ are upper and lower bounds of the $i$-th asset price respectively, and $D\coloneqq(l_1,u_1)\times\cdots\times(l_d,u_d)$.
$V_i^{\mathrm{UB}}, V_i^{\mathrm{LB}}$ are upper and lower boundary conditions for the $i$-th asset.
The boundary conditions in some typical cases of the payoff function and the payoff condition are as follows.
\begin{itemize}
\item[1.] If an \textit{up and out barrier} is set on the $i$-th asset, the payoff is zero if the asset price $S_i(t)$ exceeds $u_i$ at least once before maturity, and then the boundary condition is
\begin{align}\label{eq: up and out barrier bc}
    V^{\mathrm{UB}}_i(t, \bm{s}_{\wedge i})=0.
\end{align}
Similarly, if an \textit{down and out barrier} is set on $i$-th asset, the payoff is zero if the asset price falls below $l_i$ at least once before maturity, and then, the boundary condition is
\begin{align}\label{eq: down and out barrier bc}
    V^{\mathrm{LB}}_i(t, \bm{s}_{\wedge i})=0.
\end{align}

\item[2.]Suppose that the payoff at maturity $T$ is given by
\begin{align}\label{eq: basket payoff}
    f_{\mathrm{pay}}(\bm{S}(T))=\max(a_0 +\sum_{i=1}^d a_i S_i(T), 0),
\end{align}
with $a_0,\dots,a_d\in\mathbb{R}$.
This is the case with many derivatives.
In this form of payoff function, upper boundary or lower boundary can be set depending on the values of $a_0,\dots,a_d$.
In some cases, if either of $\left\{S_i(t)\right\}_{i\in[d]}$ is sufficiently high or low at some time $t\in(0,T)$, the payoff at $T$ is highly likely to be positive.
For example, in the case of the basket call option, that is, $a_0<0, a_1,\dots,a_d>0$, if $\bm{S}(t)=\bm{s}$ such that $s_i\gg -a_0/a_i$ for some $i\in[d]$, $f_{\mathrm{pay}}(\bm{S}(T))$ is likely to be positive.
In this situation, the derivative price is approximately equal to $E_Q\left[e^{-r(T-t)}\left(a_0 +\sum_{i=1}^d a_i S_i(T)\right)|\bm{S}(t)=\bm{s}\right]=e^{-r(T-t)}a_0 +\sum_{i=1}^d a_i s_i$. Thus, we can set
\begin{align}\label{eq: basket bc}
    V^{\mathrm{UB}}_i(t,\bm{s}_{\wedge i})=e^{-r(T-t)}a_0 +\sum_{1\leq j\leq d, j\neq i} a_j s_j + a_iu_i,
\end{align}
for sufficiently large $u_i$.
In some other cases, e.g. when $a_i<0$ and $a_j>0$ for $j\ne i$, we can set 
\begin{align}
    V^{\mathrm{LB}}_i(t, \bm{s}_{\wedge i})=e^{-r(T-t)}a_0 +\sum_{1\leq j\leq d, j\neq i} a_j s_j + a_il_i,
\end{align}
for sufficiently small $l_i$.
\end{itemize}

\subsection{Finite difference method for the BSPDE}\label{subsec:FDM}
Consider solving Eq.~\eqref{eq:bspde} using the FDM.
In the FDM, we discretize the PDE with respect to the underlying asset prices and obtain the ODE.
Then, we can use a numerical solver for ODEs, such as the Euler method, Runge-Kutta method, etc~\cite{william2007numerical}.
Note that the BSPDE is often simplified by log-transforming the asset prices as in \cite{miyamoto2022pricing}.
However, it is more convenient not to perform a log-transformation to solve the BSPDE by VQS.
This is because our formulation presented in Sec. \ref{sec:method} can only handle linear boundary conditions with respect to $s_i$ as shown in Appendix~\ref{app: decomposition}, but a logarithmic transformation will result in the terms like $e^{s_i}$.
Thus, we do not perform the log-transformation in this work.

First, the value range of each underlying asset price $s_i$ is split into $n_{\mathrm{gr}}$ grids.
That is, we take
\begin{align}
    \bm{x}^{(k)}&=\left(x_1^{(k_1)}, \dots, x_d^{(k_d)}\right)^\top,\\
    x_{i}^{\left(k_{i}\right)} &\coloneqq l_{i}+\left(k_{i}+1\right) h_{i}, \\
    k&=\sum_{i=1}^{d}n_{\mathrm{gr}}^{d-i}k_i+1,\\
    k_{i} &\coloneqq 0, \dots, n_{\mathrm{gr}}-1,\\
    h_{i} &\coloneqq\frac{u_{i}-l_{i}}{n_{\mathrm{gr}}+1}.
\end{align}
for $i \in [d]$. By this discretization, we approximate $V(t,\bm{s})$ by a vector
\begin{align}
    \bm{V}(t) \coloneqq \left(V(t, \bm{x}^{(1)}),V(t, \bm{x}^{(2)}),\dots,V(t,\bm{x}^{(N_\mathrm{gr})})\right)^\top,
\end{align}
where $N_{\mathrm{gr}}=n_{\mathrm{gr}}^d$.
We also replace the differentials by differences as,
\begin{align}
    \frac{\partial V(t, \bm{x}^{(k)})}{\partial s_i} &\rightarrow \frac{V(t, \bm{x}^{(k)}+h_i\bm{e}_i) -V(t, \bm{x}^{(k)}-h_i\bm{e}_i)}{2h_i},\label{eq: 1st diff}\\
    \frac{\partial^2 V(t, \bm{x}^{(k)})}{\partial s_i^2} & \rightarrow \frac{1}{h_i^2}\left(V(t, \bm{x}^{(k)}+h_i\bm{e}_i)+V(t, \bm{x}^{(k)}-h_i\bm{e}_i)\right.\nonumber\\
    &\left.-V(t, \bm{x}^{(k)})\right) \label{eq: 2nd diff}\\
    \frac{\partial^2 V(t, \bm{x}^{(k)})}{\partial s_i\partial s_j} &\rightarrow \frac{1}{4h_ih_j}\left(V(t, \bm{x}^{(k)}+h_i\bm{e}_i+h_j\bm{e}_j) \right.\nonumber\\
    &+V(t, \bm{x}^{(k)}-h_i\bm{e}_i-h_j\bm{e}_j)\nonumber\\
    &-V(t, \bm{x}^{(k)}-h_i\bm{e}_i+h_j\bm{e}_j)\nonumber\\
    &\left. -V(t, \bm{x}^{(k)}-h_i\bm{e}_i+h_j\bm{e}_j)\right),\label{eq: 2nd diff cross}
\end{align}
where $\bm{e}_i=(\underbrace{0, \dots, 0}_{i-1}, 1, \underbrace{0, \dots, 0}_{d-i})^\top, i\in[d]$ is a unit vector of the $i$-direction.
Introducing $\bar{V}(\tau, \bm{x}^{(i)})\coloneqq V(T-t, \bm{x}^{(i)}), i\in[d]$ and $\bar{\bm{V}}(\tau)\coloneqq\left(\bar{V}(\tau, \bm{x}^{(1)}),\bar{V}(\tau, \bm{x}^{(2)}),\dots,\bar{V}(\tau,\bm{x}^{(N_\mathrm{gr})})\right)^\top$, we eventually obtain the ODE 
\begin{align}\label{eq: discretized de}
    \frac{d}{d\tau} \bar{\bm{V}}(\tau) = F \bar{\bm{V}}(\tau) + \bm{C}(\tau)
\end{align}
and initial condition 
\begin{align}
    \bar{\bm{V}}(0)=\left(f_{\mathrm{pay}}(\bm{x}^{(1)}), \dots, f_{\mathrm{pay}}(\bm{x}^{(N_{\mathrm{gr}})})\right)^\top.
\end{align}
Here, $F$ is an $N_{\mathrm{gr}}\times N_{\mathrm{gr}}$ real matrix,
\begin{align}
    F &\coloneqq F^{\mathrm{1st}} + F^{\mathrm{2nd}} - rI^{\otimes d} \label{eq: def F}\\
    F^{\mathrm{2nd}} &\coloneqq \sum_{i=1}^d \frac{\sigma_i^2}{2h_i^2}I^{\otimes i-1}\otimes D^{\mathrm{2nd}}_{x_i} \otimes I^{\otimes d-i} \nonumber\\
    & + \sum_{i=1}^{d-1}\sum_{j=i+1}^{d}\frac{\sigma_i \sigma_j \rho_{ij}}{4h_ih_j}\nonumber \\
    & \times I^{\otimes i-1}\otimes D^{\mathrm{1st}}_{x_i} \otimes I^{\otimes j-i-1} \otimes D^{\mathrm{1st}}_{x_j} \otimes I^{\otimes d-j} \label{eq:F2nd}\\
    F^{\mathrm{1st}} &\coloneqq r\sum_{i=1}^d \frac{1}{2h_i}I^{\otimes i-1}\otimes D^{\mathrm{1st}}_{x_i} \otimes I^{\otimes d-i}, \label{eq:F1st}
\end{align}
where $I$ is a $n_{\mathrm{gr}} \times n_{\mathrm{gr}}$ identity matrix, $D_{x_i}^{1\mathrm{st}}, D_{x_i}^{2\mathrm{nd}}$ are $n_{\mathrm{gr}} \times n_{\mathrm{gr}}$ real matrices.
$\bm{C}(\tau)$ is a vector corresponding to the boundary conditions.
The elements of the $D_{x_i}^{1\mathrm{st}}, D_{x_i}^{2\mathrm{nd}}$, and $\bm{C}(\tau)$ are shown in Appendix~\ref{app: matrix elements}. 
$n_{\mathrm{gr}}$ has to be proportional to $O(\epsilon^{-1/2})$ to obtain the present price of the derivative within the accuracy $\epsilon$~\cite{miyamoto2022pricing}. 
Then, the dimension of $\bar{\bm{V}}(\tau)$ is $O((1/\epsilon)^{d/2})$.
Thus, it becomes difficult to solve the BSPDE discretized by FDM using the classical algorithm when multiple assets need to be considered.

\subsection{Approximation of the present derivative price by the expected value of the derivative price at the future time}\label{sec: approximation}
As shown in Ref.~\cite{miyamoto2022pricing}, we can evaluate the present price of the derivative by the expected value of the price at a future time $t_{\mathrm{ter}}$.
Here, we briefly review the method.
To calculate the present value of the derivative, recalling the fact that the derivative price is a martingale~\cite{shreve2004stochastic}, we evaluate $V(0,\bm{s}_0)$ as
\begin{align}\label{eq: integral}
V(0,\bm{s}_0)=e^{-rt_{\mathrm{ter}}}\int_{\mathbb{R}_+^d}d\bm{s}p(t_{\mathrm{ter}}, \bm{s})p_{\mathrm{NB}}(t_{\mathrm{ter}}, \bm{s})V(t_{\mathrm{ter}},\bm{s}),
\end{align}
where $t_{\mathrm{ter}}$ is any value in $[0,T]$, $p(t, \bm{s})$ is the probability density function of $\bm{S}(t)$, $p_{\mathrm{NB}}(t, \bm{s})$ is the probability that no event which leads to extinction of the payoff (hereafter, the {\it out event}) happens by $t$ given $\bm{S}(t_{\mathrm{ter}})=\bm{s}$, and $V(t_{\mathrm{ter}},\bm{s})$ is the derivative price at $t_{\mathrm{ter}}$ when $\bm{S}(t_{\mathrm{ter}})=\bm{s}$ and no out event happens by $t_{\mathrm{ter}}$.

Some cares must be taken to utilize Eq. \eqref{eq: integral}.
First, although we can obtain the solution of Eq.~\eqref{eq:bspde} only within the boundaries, Eq.~\eqref{eq: integral} contains the information of the events outside the boundaries.
Second, it is difficult to calculate $p_{\mathrm{NB}}(t_{\mathrm{ter}}, \bm{s})$ explicitly in the multi-asset case.
The first problem can be neglected for small $t_{\mathrm{ter}}$ since the distribution of $\bm{S}(t_{\mathrm{ter}})$ outside the boundary is negligible in this case, and so is the contribution from the outside of $D$ in Eq.~\eqref{eq: integral}.
The second problem is also solved by using sufficiently small $t_{\mathrm{ter}}$; in this case, the probability that the underlying asset prices reach any boundaries is negligible, and thus, $p_{\mathrm{NB}}(t_{\mathrm{ter}}, \bm{s})$ is nearly equal to $1$, since we are now assuming that the payoff will be paid as far as the underlying asset prices stay in the boundaries.
Therefore, for such $t_{\mathrm{ter}}$, we can evaluate $V(0, \bm{s}_0)$ as
\begin{align}\label{eq: approximate integral}
    V(0,\bm{s}_0) \simeq e^{-rt_{\mathrm{ter}}}\int_{D}d\bm{s}p(t_{\mathrm{ter}}, \bm{s})V(t_{\mathrm{ter}},\bm{s}).
\end{align}

When we use a quantum algorithm to calculate Eq.~\eqref{eq: integral} by $ \int_{D}d\bm{s}p(t_{\mathrm{ter}}, \bm{s})V(t_{\mathrm{ter}},\bm{s})\simeq \braket{\bm{p}(t_\mathrm{ter})}{\bm{V}(t_\mathrm{ter})}$, the overlap between $\ket{\bm{p}(t_{\mathrm{ter}})}$ and $\ket{\bm{V}(t_{\mathrm{ter}})}$ should be as large as possible since the number of measurements for the evaluation of the inner product decreases as the overlap become large (see Sec.~\ref{subsec: measurement on SWAP test} for details).
As the probability density function $p(t, \bm{s})$ broadens over time, taking a large $t_{\mathrm{ter}}$ results in a large overlap. 
Thus, we want to take $t_{\mathrm{ter}}$ as large as possible from this viewpoint.

Taking into account this trade-off, we set $t_{\mathrm{ter}}$ as large as possible to the extent that Eq.~\eqref{eq: integral} is well approximated with Eq.~\eqref{eq: approximate integral}.
As a conclusion, for sufficiently small $\epsilon$, we may set 
\begin{align}\label{eq: t_ter}
    t_{\mathrm{ter}} = \min &\left\{\frac{2\left(\log\left(\frac{u_1}{s_{1,0}}\right)\right)^2}{25\sigma_1^2\log\left(\frac{2\tilde{A}d(d+1)}{\epsilon}\right)},  \dots, \frac{2\left(\log\left(\frac{u_d}{s_{d,0}}\right)\right)^2}{25\sigma_d^2\log\left(\frac{2\tilde{A}d(d+1)}{\epsilon}\right)} \right. \nonumber\\
    & \left.\frac{2\left(\log\left(\frac{s_{1,0}}{l_1}\right)\right)^2}{25\sigma_1^2\log\left(\frac{2\tilde{A}d(d+1)}{\epsilon}\right)}, \dots, \frac{2\left(\log\left(\frac{s_{d,0}}{l_d}\right)\right)^2}{25\sigma_d^2\log\left(\frac{2\tilde{A}d(d+1)}{\epsilon}\right)} \right\},
\end{align}
for the approximation Eq.~\eqref{eq: approximate integral} with $O(\epsilon)$ accuracy.
Here, we assume that there exist positive constants $A_0, A_1, \dots, A_d$ such that $f_{\mathrm{pay}}$ satisfies $f_{\mathrm{pay}}(\bm{s})\leq \sum_{i=1}^d A_is_i + A_0$ for any $\bm{s}\in D$, and define $\tilde{A}=\max\left\{A_1\sqrt{u_1s_{1,0}},\dots,A_d\sqrt{u_ds_{d,0}},A_0\right\}$.
For the full detail, see Sec. 4 in \cite{miyamoto2022pricing}.

\subsection{Variational quantum simulation}\label{subsection: VQS}
In this subsection, we introduce the VQS, which is a variational quantum algorithm to solve linear ODEs~\cite{endo2020variational,yuan2019theoryofvariational,cerezo2021variational}.
Consider solving the following linear ODE,
\begin{align}
    \frac{d}{dt}\bm{v}(t) = L(t)\bm{v}(t) + \bm{u}(t), \bm{v}(0)=\bm{v}_0.
\end{align}
where $\bm{v}(t)=(v_1(t), \dots, v_{N_v}(t)), \bm{v}_0=(v_{0,1}, \dots, v_{0,N_v}), \bm{u}(t)=(u_1(t), \dots, u_{N_v}(t))\in\mathbb{C}^{N_v}$, and $L(t)$ is an (possibly non-hermitian) operator.
To simulate the vector $\bm{v}(t)$, we instead simulate an unnormalized quantum state $\ket{\bm{v}(t)}$, which is the solution of
\begin{align}\label{eq:inhomogeneous-ODE}
    \frac{d}{dt}\ket{\bm{v}(t)} = L(t)\ket{\bm{v}(t)} + \ket{\bm{u}(t)}, \ket{\bm{v}(0)}=\ket{\bm{v}_0}.
\end{align}
Here, we make three assumptions.
First, $L(t)$ can be decomposed as
\begin{align}\label{eq: decomposition of L}
    L(t) = \sum_{k=1}^{N_L}\lambda_k(t)U^L_k(t),
\end{align}
where $\lambda_k(t)$ is real, and $U^{L}_k(t)$ are quantum gates.
Second, $\ket{\bm{u}(t)}$ can be written as
\begin{align}\label{eq: decomposition of phi}
    \ket{\bm{u}(t)}=\sum_{l=1}^{N_{u}}\eta_l(t)U_l^{u}(t)\ket{0},
\end{align}
where $\eta_l(t)$ is real, and $U_l^{u}(t)$ are quantum gates.
Third, there are some constant $\alpha_{v}\in\mathbb{C}$ and an quantum gate $U_{v}$ such that $\ket{\bm{v}_0} = \alpha_{v} U_{v}\ket{0}$.
In VQS, we approximate $\ket{\bm{v}(t)}$ by an unnormalized ansatz state $\ket{\tilde{v}(\bm{\theta}(t))} \coloneqq \theta_0(t)R_1(\theta_1(t))R_2(\theta_2(t))\cdots R_{N_a}(\theta_{N_a}(t))\ket{\bm{v}_0}$ and determine parameters $\bm{\theta}(t)=(\theta_0(t),\theta_1(t),\dots,\theta_{N_a}(t))^\top\in\mathbb{R}^{N_a+1}$ by the variational principle.
Here, $R_k(\theta_k)=W_ke^{i\theta_k G_k}$ are parameterized quantum circuits, $W_k$ are quantum gates, and $G_k \in \{X, Y, Z, I\}^{\otimes n}$ are multi-qubit Pauli gates with $n$-qubit system.
By McLachlan's variational principle~\cite{McLachlan1964variational}
\begin{align}
    \min_{\bm{\theta}}\left\| \frac{d}{dt}\ket{\tilde{v}(\bm{\theta}(t))} - L(t)\ket{\tilde{v}(\bm{\theta}(t))} - \ket{\bm{u}(t)}\right\|^2,
\end{align}
we obtain the differential equation~\cite{endo2020variational}
\begin{align}\label{eq:Euler-Lagrange}
    \sum_{n=0}^{N_a} \mathcal{M}_{m,n} \dot{\theta}_n(t) = \mathcal{V}_m,
\end{align}
where 
\begin{align}
    \mathcal{M}_{i,j} &= \Re \left( \frac{\partial\bra{\tilde{v}(\bm{\theta}(t))}}{\partial\theta_i} \frac{\partial \ket{\tilde{v}(\bm{\theta}(t))}}{\partial \theta_j} \right),\label{eq:M}\\
    \mathcal{V}_j &= \sum_{k=1}^{N_L} \lambda_k(t) \Re \left( \frac{\partial\bra{\tilde{v}(\bm{\theta}(t))}}{\partial\theta_j} U^{L}_k(t) \ket{\tilde{v}(\bm{\theta}(t))} \right)\nonumber\\
    &+\sum_{l=1}^{N_{u}}\eta_l(t)\Re\left(\frac{\partial\bra{\tilde{v}(\bm{\theta}(t))}}{\partial\theta_n}U^{u}_l(t)\ket{0}\right)\label{eq:V}.
\end{align}
We can evaluate each term in Eqs.~\eqref{eq:M}\eqref{eq:V} by quantum circuits presented in Appendix \ref{app: evaluation of M, V}.
Then, we solve Eq.~\eqref{eq:Euler-Lagrange} classically and obtain $\dot{\theta}_j(t)$.
Note that the number of measurements needed to evaluate $\mathcal{M}_{i,j}$ and $ \mathcal{V}_i$  by the Hadamard test within the accuracy $\bar{\epsilon}$ is $O(|\theta_0(t)|^2/\bar{\epsilon}^2)$.
This is because $\mathcal{M}_{i,j}$ and $ \mathcal{V}_i$ contain the normalization factor $\theta_0(t)$ when $i>0$ or $j>0$ (see Appendix \ref{app: evaluation of M, V}).
We assume that $|\theta_0(t)|$ is upper-bounded by some constant.
In derivative pricing, $|\theta_0(t)|^2$ is about a ratio of the sum of the squares of the derivative prices at time $T-t$ to the sum of the squares of the payoff function at maturity.
Since the derivative price is the expected value of the payoff function, this assumption is satisfied if the value range of the payoff function is finite.
Starting from $t=0$, we obtain the time evolution of $\bm{\theta}(t)$ by repeating
\begin{align}
    \bm{\theta}(t+\Delta t) \leftarrow \bm{\theta}(t) + \dot{\bm{\theta}}(t)\Delta t,
\end{align} where $\Delta t$ is an interval in time direction.
Consequently, we obtain $\ket{\tilde{v}(\bm{\theta}(t))}$ which approximates $\ket{\bm{v}(t)}$.

\section{Proposed method}\label{sec:method}
In this section, we describe the variational quantum algorithm for derivative pricing and the computational complexity of the proposed method.
The overall algorithm is shown in Algo.~\ref{algo: main}.
We assume that $n_{\mathrm{gr}}^d=2^n$ with the $n$-qubit system.
\begin{algorithm}[H]
    \caption{Derivative Pricing with Variational Quantum Algorithms}
    \label{algo: main}
    \begin{algorithmic}[1]
    \State Prepare $ a_{p}U_{p}$ such that $a_{p}U_{p}\ket{0}=\ket{\psi_{p}} \simeq\sum_{k=1}^{N_{\mathrm{gr}}}p_k(t_{\mathrm{ter}})\ket{k}$ by VQS for Fokker-Planck equation or quantum generative models.
    \State Prepare $\alpha_{V} U_{V}$ such that $\alpha_{V} U_{V}\ket{0} = \ket{\psi_{V}} \simeq \sum_{k=1}^{N_\mathrm{gr}}f_{\mathrm{pay}}(\bm{x}^{(k)})\ket{k}$ by quantum generative models.
    \State Calculate $\ket{\tilde{v}(\bm{\theta}(\tau_{\mathrm{ter}}))}$ by performing VQS from $\tau=0$ to $\tau=\tau_{\mathrm{ter}}$.
    \State Perform the SWAP test and get an estimation of $\left|\braket{\psi^{p}}{\tilde{v}(\bm{\theta}(\tau_{\mathrm{ter}}))}\right|^2$
    \State $V_0 \gets e^{-rt_{\mathrm{ter}}}\braket{\psi_{p}}{\tilde{v}(\bm{\theta}(\tau_{\mathrm{ter}}))}$.
    \end{algorithmic}
\end{algorithm}

First, we set $\tau_{\mathrm{ter}}=T-t_{\mathrm{ter}}$, where $t_{\mathrm{ter}}$ is defined in Eq.~\eqref{eq: t_ter}.
We also set $N_{\tau}$, which is the number of steps for VQS.
To perform VQS, we need to represent the operator corresponding to $F$ in Eq.~\eqref{eq: def F} and the operator $\tilde{G}$ such that $\tilde{G}\ket{0}=\ket{\bm{C}}=\sum_{k=1}^{N_{\mathrm{gr}}}C_k(\tau)\ket{k}$ by a linear combination of quantum gates, respectively, because of the assumptions Eqs.~\eqref{eq: decomposition of L} and \eqref{eq: decomposition of phi}.
Such decomposition can be obtained in a similar way to Ref.~\cite{kubo2021variational,alghassi2022variational} and is shown in Appendix~\ref{app: decomposition}.
$F$ can be represented as a sum of $O(d^2n^4)$ unitaries each of which requires at most $O(n^2)$ gates to be implemented.
$\tilde{G}$ for typical boundary conditions discussed in Sec.~\ref{subsec:FDM} can be represented by $O(d^3n^2)$ unitaries, which require at most $O(n^2)$ gates to be implemented.


Second, we prepare the unnormalized state
\begin{align}\label{eq: approximate p}
    \ket{\psi_{p}}&\coloneqq\alpha_{p}U_{p}\ket{0}\nonumber\\
    &\simeq\ket{\bm{p}(t_{\mathrm{ter}})}\nonumber\\
    &=\sum_{k=1}^{N_{\mathrm{gr}}}p_k(t_{\mathrm{ter}})\ket{k},
\end{align}
where $\alpha_{p}\in\mathbb{C}$, and $U_{p}$ is an quantum gate.
$p_k(t_{\mathrm{ter}})$ is a probability that the underlying asset prices is on $\bm{x}^{(k)}$ at $t_{\mathrm{ter}}$.
We can obtain such $\alpha_{p}$ and $U_{p}$ by solving the Fokker-Planck equation, which describes the time evolution of the probability density function, using VQS~\cite{kubo2021variational, alghassi2022variational}.
Alternatively, they can also be obtained by quantum generative models ~\cite{zoufal2019quantum,situ2020quantum,lloyd2018quantum, dallaire2018quantum,kyriienko2022protocols} since the probability density function of the underlying asset price at any $t\in[0,T]$ can be obtained analytically under the BS model (see Eq.~\eqref{eq: lognormal pdf} in Sec.~\ref{subsec: measurement on SWAP test}).

Third, we prepare $\alpha_{V}\in\mathbb{C}$ and $U_{V}$ such that $\alpha_{V}U_{V}\ket{0}\eqqcolon\ket{\psi_{V}}$ approximates the initial state of the discretized BSPDE, that is, 
\begin{align}
    \ket{\psi_{V}}&\simeq \ket{\bar{\bm{V}}(0)}\nonumber\\
    &= \sum_{k=1}^{N_\mathrm{gr}}f_{\mathrm{pay}}(\bm{x}^{(k)})\ket{k}
\end{align}
To find such $\alpha_{V}$ and $U_{V}$, we can use the quantum generative models ~\cite{zoufal2019quantum,situ2020quantum,lloyd2018quantum, dallaire2018quantum,kyriienko2022protocols}.

Fourth, we solve the BSPDE from $\tau=0$ to $\tau_{\mathrm{ter}}$ using VQS and obtain an unnormalized state
\begin{align}\label{eq: approximate V}
    \ket{\tilde{v}(\bm{\theta}(\tau_{\mathrm{ter}}))}\simeq\ket{\bar{\bm{V}}(\tau_{\mathrm{ter}})}=\sum_{k=1}^{N_{\mathrm{gr}}} \bar{\bm{V}}_k(\tau_{\mathrm{ter}}, \bm{x}^{(k)}) \ket{k}
\end{align}
where 
\begin{align}
    \ket{\tilde{v}(\bm{\theta}(\tau_{\mathrm{ter}}))}&\coloneqq\theta_0(\tau_{\mathrm{ter}})R_1(\theta_1(\tau_{\mathrm{ter}}))R_2(\theta_2(\tau_{\mathrm{ter}}))\nonumber\\
    &\cdots  R_{N_{a}}(\theta_{N_{a}}(\tau_{\mathrm{ter}}))\ket{\psi_{V}}
\end{align}
$\{R_k\}_{k\in [N_a]}$ are parameterized quantum circuits, and $\bm{\theta}(\tau_{\mathrm{ter}})\coloneqq(\theta_0(\tau_{\mathrm{ter}}),\dots,\theta_{N_a}(\tau_{\mathrm{ter}}))^\top\in\mathbb{R}^{N_a+1}$ is the variational parameters.
Note that $\theta_0(0)R_1(\theta_1(0))R_2(\theta_2(0))R_{N_{a}}(\theta_{N_{a}}(0))$ should be an identity operator to satisfy $\ket{\tilde{v}(\bm{\theta}(0))} \simeq \ket{\bar{\bm{V}}(0)}$.
For example, the ansatz shown in Fig.~\ref{fig: ansatz} in Sec.~\ref{sec: numerical results} with even number of layers can be used as $\{R_k\}_{k\in [N_a]}$ that satisfies this condition with the parameters $\bm{\theta}(0)=(0,\dots,0)^\top$ since $RY$ gates are identity for the parameters, and $CZ$ layers cancel each other and also become identity.

Finally, we use the SWAP test~\cite{carlos2013swap} for two normalized states $U_{p}\ket{0}, R_1(\theta_1(\tau_{\mathrm{ter}})) \cdots  R_{N_a}(\theta_{N_a}(\tau_{\mathrm{ter}}))U_{V}\ket{0}$ and obtain
\begin{align}
    &\left|\braket{\psi_{p}}{\tilde{v}(\bm{\theta}(\tau_{\mathrm{ter}}))}\right|^2\nonumber\\
    &=\left|\alpha_p\alpha_{V}\theta_0(\tau_{\mathrm{ter}})\right|^2\nonumber\\
    &\times\left|\bra{0}U_p^{\dagger} R_1(\theta_1(\tau_{\mathrm{ter}})) \cdots  R_{N_a}(\theta_{N_a}(\tau_{\mathrm{ter}}))U_{V}\ket{0}\right|^2.
\end{align}
As discussed in Sec.~\ref{sec: approximation}, the present price of the derivative is approximated by the inner product $\braket{\bm{p}(t_{\mathrm{ter}})}{\bar{\bm{V}}(\tau_{\mathrm{ter}})}$ discounted by the risk-free rate.
We can approximate the inner product by the square root of the result of the SWAP test and obtain the present price of the derivative by
\begin{align}\label{eq: evaluation of V0}
    V_0 \simeq e^{-rt_{\mathrm{ter}}}\braket{\psi_{p}}{\tilde{v}(\bm{\theta}(\tau_{\mathrm{ter}}))}.
\end{align}

For the third and fourth parts, we may take a slightly different approach.
That is, we find $\bm{\theta}(0)$ such that
\begin{align}
    \ket{\bar{\bm{V}}(0)}\simeq\theta_0(0)R_1(\theta_1(0))R_2(\theta_2(0))\cdots R_{N_a}(\theta_{N_a}(0))\ket{0}
\end{align}
and obtain 
\begin{align}
    \ket{\tilde{v}(\bm{\theta}(\tau_{\mathrm{ter}}))}&=\theta_0(\tau_{\mathrm{ter}})R_1(\theta_1(\tau_{\mathrm{ter}}))R_2(\theta_2(\tau_{\mathrm{ter}}))\nonumber\\
    &\cdots R_{N_a}(\theta_{N_a}(\tau_{\mathrm{ter}}))\ket{0}\nonumber\\
    &\simeq\ket{\bar{\bm{V}}(\tau_{\mathrm{ter}})}
\end{align}
using VQS.
This approach may reduce the number of gates by eliminating $U_V$, but since the ansatz for the initial state also serves as the ansatz for VQS, the number of gates required for the ansatz may become larger.
For this reason, it is difficult to say which approach is better in general, but we adopt the one in Algo.~\ref{algo: main} for the numerical simulation in Sec.~\ref{sec: numerical results}.

\subsection{The number of measurements in the SWAP test}\label{subsec: measurement on SWAP test}
In this subsection, we estimate the number of measurements required for the SWAP test.
For simplicity, we consider the case where $\ket{\tilde{v}(\bm{\theta}(\tau_{\mathrm{ter}}))}=\ket{\bar{\bm{V}}(\tau_{\mathrm{ter}})}$ and $\ket{\psi_{p}}=\ket{\bm{p}(t_{\mathrm{ter}})}$.
We perform the SWAP test for two normalized states $\ket{\tilde{\bm{p}}}$ and $\ket{\tilde{\bar{\bm{V}}}}$ such that $\ket{\bm{p}(t_{\mathrm{ter}})} = \alpha \ket{\tilde{\bm{p}}}, \ket{\bar{\bm{V}}(\tau_{\mathrm{ter}})} = \beta \ket{\tilde{\bar{\bm{V}}}}$, where 
\begin{align}
    \alpha &= \sqrt{\sum_{k=1}^{N_{\mathrm{gr}}}p_k(t_\mathrm{ter})^2},\\
    \beta &= \sqrt{\sum_{k=1}^{N_{\mathrm{gr}}}\bar{V}(\tau_\mathrm{ter}, \bm{x}^{(k)})^2}.
\end{align}
To obtain the value of the inner product $\left|\innerproduct{\tilde{\bm{p}}}{\tilde{\bar{\bm{V}}}}\right|^2$ with precision $\bar{\varepsilon}$, the SWAP test requires $O(\frac{1}{\bar{\varepsilon}^2})$ measurements~\cite{carlos2013swap}.
When we have the estimation $\widetilde{\left|\innerproduct{\tilde{\bm{p}}}{\tilde{\bar{\bm{V}}}}\right|^2}$ such that
\begin{align}
    \left|\left|\innerproduct{\tilde{\bm{p}}}{\tilde{\bar{\bm{V}}}}\right|^2-\widetilde{\left|\innerproduct{\tilde{\bm{p}}}{\tilde{\bar{\bm{V}}}}\right|^2}\right| < \bar{\varepsilon},
\end{align}
the estimation of the inner product of unnormalized states $\widetilde{\left|\innerproduct{\bm{p}(t_{\mathrm{ter}})}{\bar{\bm{V}}(\tau_{\mathrm{ter}})}\right|^2}$ satisfies
\begin{align}
    \left|\left|\innerproduct{\bm{p}(t_{\mathrm{ter}})}{\bar{\bm{V}}(\tau_{\mathrm{ter}})}\right|^2-\widetilde{\left|\innerproduct{\bm{p}(t_{\mathrm{ter}})}{\bar{\bm{V}}(\tau_{\mathrm{ter}})}\right|^2}\right| < \alpha^2\beta^2\bar{\varepsilon}.
\end{align}
Thus, $O(\frac{\alpha^4\beta^4}{\varepsilon^2})$ measurements are required to obtain $\left|\innerproduct{\bm{p}(t_{\mathrm{ter}})}{\bar{\bm{V}}(\tau_{\mathrm{ter}})}\right|^2$ with precision $\varepsilon\coloneqq\alpha^2\beta^2\bar{\varepsilon}$.
Note that since we can classically calculate $\alpha$ by the analytical form of $p(t,\bm{s})$, and $\beta$ is calculated by $\alpha_{V}\theta_0(\tau_{\mathrm{ter}})$, we can determine the number of measurements before the SWAP test from VQS results.

To estimate the number of measurements of the SWAP test, we estimate $\alpha^2\beta^2$, which is calculated as
\begin{align}
    \alpha^2\beta^2&=\left(\sum_{k=1}^{N_{\mathrm{gr}}}p_k(t_\mathrm{ter})^2\right)\left(\sum_{k=1}^{N_{\mathrm{gr}}}\bar{V}(\tau_\mathrm{ter}, \bm{x}^{(k)})^2\right)\nonumber\\
    &=\left(\sum_{k=1}^{N_{\mathrm{gr}}}p_k(t_\mathrm{ter})^2\right)\left(\sum_{k=1}^{N_{\mathrm{gr}}}f_{\mathrm{pay}}(\bm{x}^{(k)})^2\right)\nonumber\\
    &\times\frac{\sum_{k=1}^{N_{\mathrm{gr}}}\bar{V}(\tau_\mathrm{ter}, \bm{x}^{(k)})^2}{\sum_{k=1}^{N_{\mathrm{gr}}}f_{\mathrm{pay}}(\bm{x}^{(k)})^2}.
\end{align}
Although it is difficult to estimate the factor $\sum_{k=1}^{N_{\mathrm{gr}}}\bar{V}(\tau_\mathrm{ter}, \bm{x}^{(k)})^2/\sum_{k=1}^{N_{\mathrm{gr}}}f_{\mathrm{pay}}(\bm{x}^{(k)})^2$ in advance, we assume that the factor is bounded by some constant $\zeta$. 
This assumption means that the rate of change in derivative prices over time is suppressed by a certain constant.
Under the assumption, we estimate $\left(\sum_{k=1}^{N_{\mathrm{gr}}}f_{\mathrm{pay}}(\bm{x}^{(k)})^2\right)\left(\sum_{k=1}^{N_{\mathrm{gr}}}p_k(t_\mathrm{ter})^2\right)$.
We assume that $f_{\mathrm{pay}}(\bm{x})$ for $\bm{x}\in D$ is upper bounded by some constant $B$.
For example, in the case of the basket call option,
\begin{align}
    f_{\mathrm{pay}}(\bm{x}) &= \max(a_0 + \sum_{i=1}^{d} a_ix_i - K, 0) \nonumber\\
    &\leq a_0 + \sum_{i=1}^{d} a_i x_i\nonumber\\
    &\leq a_0 + \sum_{i=1}^{d} a_i u_i
\end{align}
holds.
From this assumption, we obtain
\begin{align}\label{eq: upper bound of fpay}
    \sum_{k=1}^{N_{\mathrm{gr}}}f_{\mathrm{pay}}(\bm{x}^{(k)})^2&\leq\sum_{k=1}^{N_{\mathrm{gr}}}B^2\nonumber\\
    &=N_{\mathrm{gr}}B^2.
\end{align}

On the other hand, the probability density function of $d$-dimensional geometric Brownian motion with $\bm{x}(0)=\bm{x}_0\coloneqq(x_{0,1},\dots,x_{0,d})^{\top}$ is
\begin{align}\label{eq: lognormal pdf}
    p(t, \bm{x}) &= \frac{1}{(2\pi t)^{d/2} \left(\prod_{i=1}^d\sigma_ix_i\right) \sqrt{\det\rho}} \nonumber\\
    & \times \exp \left(-\frac{1}{2}(\ln\bm{x}-\bm{\mu})^\top \Sigma^{-1}(\ln\bm{x}-\bm{\mu})\right),
\end{align}
where
\begin{align}
    \bm{\mu} = \left(\left(r-\frac{\sigma_1^2}{2}\right)t-x_{0,1}, \dots, \left(r-\frac{\sigma_d^2}{2}\right)t-x_{0,d} \right)^\top.
\end{align}
The square of probability density function is
\begin{align}
    p(t, \bm{x})^2 &= \left(\frac{1}{(2\pi t)^{d/2} \left(\prod_{i=1}^d\sigma_ix_i\right) \sqrt{\det\rho}}\right)^2 \nonumber\\
    & \times \exp \left(-(\ln\bm{x}-\bm{\mu})^\top \Sigma^{-1}(\ln\bm{x}-\bm{\mu})\right) \nonumber\\
    &=\frac{\gamma(t)}{\prod_{i=1}^{d}x_i} \frac{1}{(2\pi t)^{d/2} \left(\prod_{i=1}^d\frac{\sigma_ix_i}{2}\right) \sqrt{\det\rho}}\nonumber\\
    & \times \exp \left(-\frac{1}{2}(\ln\bm{x}-\bm{\mu})^\top \left(\frac{1}{2}\Sigma\right)^{-1}(\ln\bm{x}-\bm{\mu})\right) \nonumber\\
    &=\frac{\gamma(t)}{\prod_{i=1}^{d}x_i}\varphi(t, \bm{x}),
\end{align}
where 
\begin{align}\label{eq: gamma}
    \gamma(t) = \frac{1}{(8\pi t)^{d/2}\prod_{i=1}^{d}\sigma_i},
\end{align}
and $\varphi(t,\bm{x})$ is a probability density function of some log-normal distribution.
Using the probability distribution function, the square sum of the discretized density function is represented by
\begin{align}\label{eq: upper bound of p}
    \sum_{k=1}^{N_{\mathrm{gr}}}p_k(t_\mathrm{ter})^2 &= \sum_{k=1}^{N_{\mathrm{gr}}} \left(p(t_\mathrm{ter},\bm{x}^{(k)})\right)^2 \left(\prod_{i=1}^d{h_i}\right)^2\nonumber\\
    &=\sum_{k=1}^{N_{\mathrm{gr}}}\frac{\gamma(t_{\mathrm{ter}})}{\prod_{i=1}^{d}x_i^{(k_i)}}\varphi(t_{\mathrm{ter}}, \bm{x}^{(k)})\left(\prod_{i=1}^d{h_i}\right)^2\nonumber\\
    &\leq\frac{\gamma(t_{\mathrm{ter}})}{\prod_{i=1}^{d}l_i}\sum_{k=1}^{N_{\mathrm{gr}}}\varphi(t_{\mathrm{ter}}, \bm{x}^{(k)})\left(\prod_{i=1}^d{h_i}\right)^2\nonumber\\
    &\simeq \frac{\gamma(t_{\mathrm{ter}})}{\prod_{i=1}^{d}l_i}\prod_{i=1}^d{h_i}\int_{\mathbb{R}_+^d}\varphi(t_{\mathrm{ter}}, \bm{x})d\bm{x}\nonumber\\
    &= \frac{\gamma(t_{\mathrm{ter}})}{\prod_{i=1}^{d}l_i}\prod_{i=1}^d{h_i}\nonumber\\
    &=\frac{\gamma(t_{\mathrm{ter}})}{\prod_{i=1}^{d}l_i}\frac{1}{N_{\mathrm{gr}}}\prod_{i=1}^d(u_i-l_i).
\end{align}
From Eqs.~\eqref{eq: upper bound of fpay} and \eqref{eq: upper bound of p}, we obtain
\begin{align}
    \alpha^2\beta^2 &\lesssim \zeta B^2 \frac{1}{(8\pi t_{\mathrm{ter}})^{d/2}}\prod_{i=1}^d\frac{1}{\sigma_i}\left(\frac{u_i}{l_i} - 1\right)\nonumber\\
    &\eqqcolon \Xi.
\end{align}
Since $t_{\mathrm{ter}}$ is lower bounded by
\begin{align}
    t_{\mathrm{ter}} &= \min \left\{\frac{2\left(\log\left(\frac{u_1}{s_{1,0}}\right)\right)^2}{25\sigma_1^2\log\left(\frac{2\tilde{A}d(d+1)}{\epsilon}\right)},  \dots, \frac{2\left(\log\left(\frac{u_d}{s_{d,0}}\right)\right)^2}{25\sigma_d^2\log\left(\frac{2\tilde{A}d(d+1)}{\epsilon}\right)} \right. \nonumber\\
    & \left.\frac{2\left(\log\left(\frac{s_{1,0}}{l_1}\right)\right)^2}{25\sigma_1^2\log\left(\frac{2\tilde{A}d(d+1)}{\epsilon}\right)}, \dots, \frac{2\left(\log\left(\frac{s_{d,0}}{l_d}\right)\right)^2}{25\sigma_d^2\log\left(\frac{2\tilde{A}d(d+1)}{\epsilon}\right)} \right\},\nonumber\\
    &\geq \frac{2 \left(\log \chi_{\mathrm{min}}\right)^2}{25\sigma_{\mathrm{max}}^2}\left(\log \frac{2\tilde{A}d(d+1)}{\epsilon}\right)^{-1},
\end{align}
where $\sigma_{\mathrm{max}}\coloneqq \max_{i\in[d]}\left\{\sigma_i\right\}$, and $\chi_{\min}\coloneqq \min_{i\in[d]}\left\{u_i/s_{i,0}\right\}\cup\left\{s_{i,0}/l_i\right\}$, we obtain
\begin{align}
    \Xi&\leq \zeta B^2 \nonumber\\
    & \times \left(\frac{5}{4\pi^2}\frac{\xi_{\mathrm{max}}-1}{\log \chi_{\mathrm{min}}}\frac{\sigma_{\mathrm{max}}}{\sigma_{\mathrm{min}}}\right)^d
    \left(\log \frac{2\tilde{A}d(d+1)}{\epsilon}\right)^{d/2},
\end{align}
where $\sigma_{\mathrm{min}}\coloneqq \min_{i\in[d]}\left\{\sigma_i\right\}$, and $\xi_{\max}\coloneqq \max_{i\in[d]}\left\{{u_i/l_i}\right\}$\footnote{When $\xi_{\mathrm{max}}$ is close to $1$, one may find it strange that as $\Xi$ decreases exponentially with respect to the number of assets $d$, and then, the number of measurements also decrease exponentially.
We show that such an exponential decrease does not occur by evaluating the lower bound of $\Xi$.
See Appendix~\ref{app: lower bound of Xi} for details.
}.
We find that the number of measurements required by the SWAP test is
\begin{align}\label{eq: number of measurements to perform SWAP test}
    N_{\mathrm{SWAP}} = \frac{\zeta^2B^4}{\varepsilon^2}  \left(\frac{5}{4\pi^2}\frac{\xi_{\mathrm{max}}-1}{\log \chi_{\mathrm{min}}}\frac{\sigma_{\mathrm{min}}}{\sigma_{\mathrm{max}}}\right)^{2d}
    \left(\log \frac{2\tilde{A}d(d+1)}{\epsilon}\right)^{d}.
\end{align}
Note that $N_{\mathrm{SWAP}}$ does not have the dependency of the form like $(1/\epsilon)^{O(d)}$, which means that the proposed method achieves a significant speedup over classical FDM with respect to $\epsilon$ and $d$, when the other parts of the proposed method are sufficiently efficient.

Here, we consider the limit of $t_{\mathrm{ter}}\rightarrow0$.
This corresponds to retrieving one amplitude of the computational basis from $\ket{\bm{V}(0)}$ as in \cite{gonzalezconde2021pricing,fontanela2021quantum}.
In this case, the probability density function (Eq.~\eqref{eq: lognormal pdf}) is a delta function, which means that the present price of the underlying assets is $\bm{x}_0$ with probability $1$.
Assuming that $\bm{x}_0$ is on a grid point with the index $k_0$, $p_k(0)$ is 1 for $k=k_0$ and 0 otherwise, and the sum of the squares of $p_k(0)$ is $1$.
As a result, $\alpha^2\beta^2$ is upper-bounded as follows,
\begin{align}
    \alpha^2\beta^2 \leq \zeta N_{\mathrm{gr}} B^2
\end{align}
Thus, the number of the measurement is proportional to $N_{\mathrm{gr}}=n_{\mathrm{gr}}^d$, and the quantum speedup will be lost.



\subsection{Computational complexity of proposed method}\label{subsec: complexity}
Here, we discuss the computational complexity of our algorithm.
We assume that the number of quantum gates required for preparing $\ket{\bm{p}(t_\mathrm{ter})}$ and $\ket{\bar{\bm{V}}(0)}$ are $N^{p}_{\mathrm{gate}}$ and $N^{V}_{\mathrm{gate}}$ respectively.
We also assume that the number of measurements required to prepare $\ket{\bm{p}(t_\mathrm{ter})}$ and $\ket{\bar{\bm{V}}(0)}$ are $N^{p}_{\mathrm{measure}}$ and $N^{V}_{\mathrm{measure}}$ respectively.
$N^{{p}}_{\mathrm{gate}}, N^{{V}}_{\mathrm{gate}}, N^{p}_{\mathrm{measure}}$, and $N^{V}_{\mathrm{measure}}$ depend on the implementation of the generative models, but we assume that all of them are $O(\mathrm{poly}(d\log (1/\epsilon)))$.
This means that we assume that the generative models efficiently generate the (unnormalized) quantum states.
Note that VQS requires controlled versions of  $U_k^{L}$, $U_l^{u}$ in Eq.~\eqref{eq:V}, or those of $\mathcal{R}U_{v}$ where $\mathcal{R}$ is defined in Eq.~\eqref{eq: R} (see Appendix~\ref{app: evaluation of M, V}).
Since $U_k^{L}$ and $U_l^{u}$ are terms of the linear combination of $F$ and $\tilde{G}$, respectively, they are made by $O(n^2)=O(d\log(1/\epsilon))$ gates.
Thus, $O(\mathrm{poly}(d\log (1/\epsilon))$ gates are required for the control unitaries of $U_k^{L}$ and $U_l^{u}$.
Assuming $\mathcal{R}U_{v}$ is made by $O(\mathrm{poly}(d\log (1/\epsilon))$ gates, the controlled-$\mathcal{R}U_{v}$ gate requires $O(\mathrm{poly}(d\log (1/\epsilon))$ gates.
Consequently, quantum circuits containing $O(\mathrm{poly}(d\log (1/\epsilon))$ quantum gates is required for VQS.
We assume that the number of measurements to estimate $\mathcal{M}_{i,j}$ and $\mathcal{V}_i$ are $N_{\mathrm{measure}}^{\mathrm{VQS}}$.
The number of quantum gates to perform the SWAP test is $O(\mathrm{poly}(d\log (1/\epsilon)))$ since the SWAP test requires $O(n)=O(d\log(1/\epsilon))$ quantum gates in addition to the quantum gates to generate $\ket{\bm{p}(t_{\mathrm{ter}})}$ and $\ket{\bar{\bm{V}}(\tau)}$~\cite{carlos2013swap}.
The number of measurements for the SWAP test is $N_{\mathrm{SWAP}}$ in Eq.~\eqref{eq: number of measurements to perform SWAP test}.
The summary of the complexities of the proposed method is shown in Table~\ref{tab: complexity}.
\begin{table}[hbtp]
\centering
\begin{tabular}{c|c|c}
    \hline
     Part of the algorithm & \# of quantum gates & \# of measurements \\
     \hline\hline
     Preparing $\ket{\bm{p}_{\mathrm{ter}}}$ & $N_{\mathrm{gate}}^{p}$ & $N_{\mathrm{measure}}^{p}$\\
      \hline
     Preparing $\ket{\bar{\bm{V}}_{\mathrm{ter}}}$ & $N_{\mathrm{gate}}^{V}$ & $N_{\mathrm{measure}}^{V}$\\
     \hline
     VQS &  
     $O(\mathrm{poly}(d\log (1/\epsilon)))$ & $N_{\mathrm{measure}}^{\mathrm{VQS}} N_{\tau}$\\
     \hline
     SWAP test & $O(\mathrm{poly}(d\log (1/\epsilon)))$ & $N_{\mathrm{SWAP}}$ in Eq.\eqref{eq: number of measurements to perform SWAP test} \\
     \hline
\end{tabular}
\caption{The complexities of the proposed method.}
\label{tab: complexity}
\end{table}

Note that, although there remains the exponential dependency with respect to $d$ in $N_{\mathrm{SWAP}}$, the time complexity does not have any factor like $(1/\epsilon)^{O(d)}$, as discussed in Sec.~\ref{subsec: measurement on SWAP test}.
This is the possible advantage of our method since the complexity of the classical FDM and conventional quantum algorithm have a factor like $(1/\epsilon)^{O(d)}$.

\section{Numerical Results}\label{sec: numerical results}
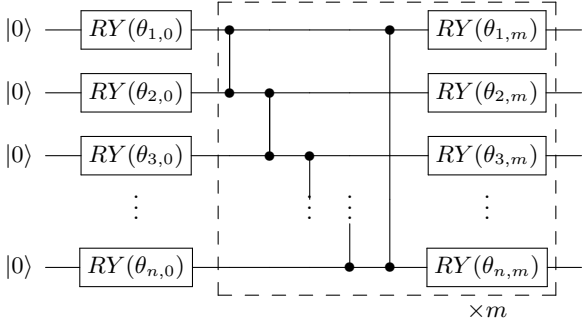
\begin{figure}[t]
    \centering
    \begin{equation*}
        \mbox{
        \Qcircuit @C=1.4em @R=1em {
        \lstick{\ket{0}} & \gate{RY(\theta_{1,0})} & \ctrl{1} & \qw        & \qw     & \qw  & \ctrl{3}   & \gate{RY(\theta_{1,m})} & \qw   \\
        \lstick{\ket{0}} & \gate{RY(\theta_{2,0})} &   \ctrl{-1}    & \ctrl{1} & \qw    & \qw   & \qw    & \gate{RY(\theta_{2,m})}  & \qw  \\
        \lstick{\ket{0}} & \gate{RY(\theta_{3,0})} &   \qw      & \ctrl{-1}    &\ctrl{1} & \qw & \qw & \gate{RY(\theta_{3,m})}  & \qw     \\
        &\vdots&&&\vdots&\vdots&&\vdots\\ 
        &&&&&&&\\
        \lstick{\ket{0}} & \gate{RY(\theta_{n,0})} &   \qw      & \qw    & \qw &   \ctrl{-1}   & \ctrl{-2} & \gate{RY(\theta_{n,m})} & \qw \gategroup{1}{3}{6}{8}{.7em}{--}   \\
        &&&&&&&\times m
        }
        }
    \end{equation*}
    \caption[ansatz circuit]{In a depth-$m$ circuit, CZ and RY gates (enclosed by dashed lines) are repeated $m$-times. The circuit has $n(m+1)$ parameters.
    }\label{fig: ansatz}
\end{figure}

In this section, we validate the proposed method using numerical calculations.
This experiment focuses on a single asset double knock-out barrier option, which contains both \textit{up and out} and \textit{down and out} conditions.
According to \cite{kunimoto1992pricing}, the analytical solution for the single asset double barrier option $ \tilde{V}$ with an upper bound $u$ and a lower bound $l$ is 
\begin{align}
    \tilde{V}(t) &= S_0 \sum_{n=-\infty}^{\infty}\left\{\left(\frac{u^n}{l^n}\right)^c\left[\mathcal{N}(d_{1n})-\mathcal{N}(d_{2n})\right]\right. \nonumber\\
    &\left. - \left(\frac{u^{n+1}}{l^nS_0}\right)^{c}\left[\mathcal{N}(d_{3n})-\mathcal{N}(d_{4n})\right]\right\} \nonumber\\
    & - K e^{-r\tau} \nonumber\\
    &\times\sum_{n=-\infty}^{\infty} \left\{\left(\frac{u^n}{l^n}\right)^{c-2}\left[\mathcal{N}(d_{1n}-\sigma\sqrt{\tau})-\mathcal{N}(d_{2n}-\sigma\sqrt{\tau})\right]\right.\nonumber\\
    &\left. -\left(\frac{u^{n+1}}{l^nS_0}\right)^{c-2}\left[\mathcal{N}(d_{3n}-\sigma\sqrt{\tau})-\mathcal{N}(d_{4n}-\sigma\sqrt{\tau})\right] \right\},
\end{align}
where
\begin{align}
d_{1n} &= \frac{\ln\left(\frac{S_0}{K}\left(\frac{u}{l}\right)^{2n}\right) - \left(r + \frac{\sigma^2}{2}\right)\tau}{\sigma\sqrt{\tau}},\\
d_{2n} &= \frac{\ln \left(S_0\frac{u^{2n-1}}{l^{2n}}\right) - \left(r + \frac{\sigma^2}{2}\right)\tau}{\sigma\sqrt{\tau}},\\
d_{3n} &= \frac{\ln\left(\frac{u^{2n+2}}{KS_0l^{2n}}\right) - \left(r + \frac{\sigma^2}{2}\right)\tau}{\sigma\sqrt{\tau}},\\
d_{4n} &= \frac{\ln\left(S_0\frac{u^{2n+1}}{l^{2n}}\right) - \left(r + \frac{\sigma^2}{2}\right)\tau}{\sigma\sqrt{\tau}},\\
c &= \frac{2r}{\sigma} + 1,
\end{align}
and $\mathcal{N}(\cdot)$ is the cumulative distribution function of the standard normal distribution.
We compare the results obtained by the proposed method with the analytical solution.
We use the Euler method for the time evolution of the parameter (Eq.~\eqref{eq:Euler-Lagrange}).
The step size for the Euler method is $\Delta\tau=2.5\times10^{-5}$.
The parameters are $r=0.001, \sigma\coloneqq\sigma_1=0.3, T=1, S_0=1, l\coloneqq l_1=0.5, u\coloneqq u_1=2.0, K=1$.
The ansatz of VQS for solving the BS model is shown in Fig.~\ref{fig: ansatz}.
This ansatz repeats $m$ parameterized layers consisting of $n$ $RY$ gates and an entanglement layer consisting of $CZ$ gates.
The ansatz have $n(m+1)$ parameters.
We do not consider noise and statistical errors in the simulation of quantum circuits.
In addition, we assume that the initial state $\ket{\bar{\bm{V}}(0)}$ and $\ket{\bm{p}(t)}$ for all $t\in[0, T]$ are given.
For the simulation of quantum states, we use NumPy~\cite{harris2020array}.

\subsection{Parameter dependencies of VQS results}
\begin{figure}
    \centering
    \includegraphics[width=\linewidth]{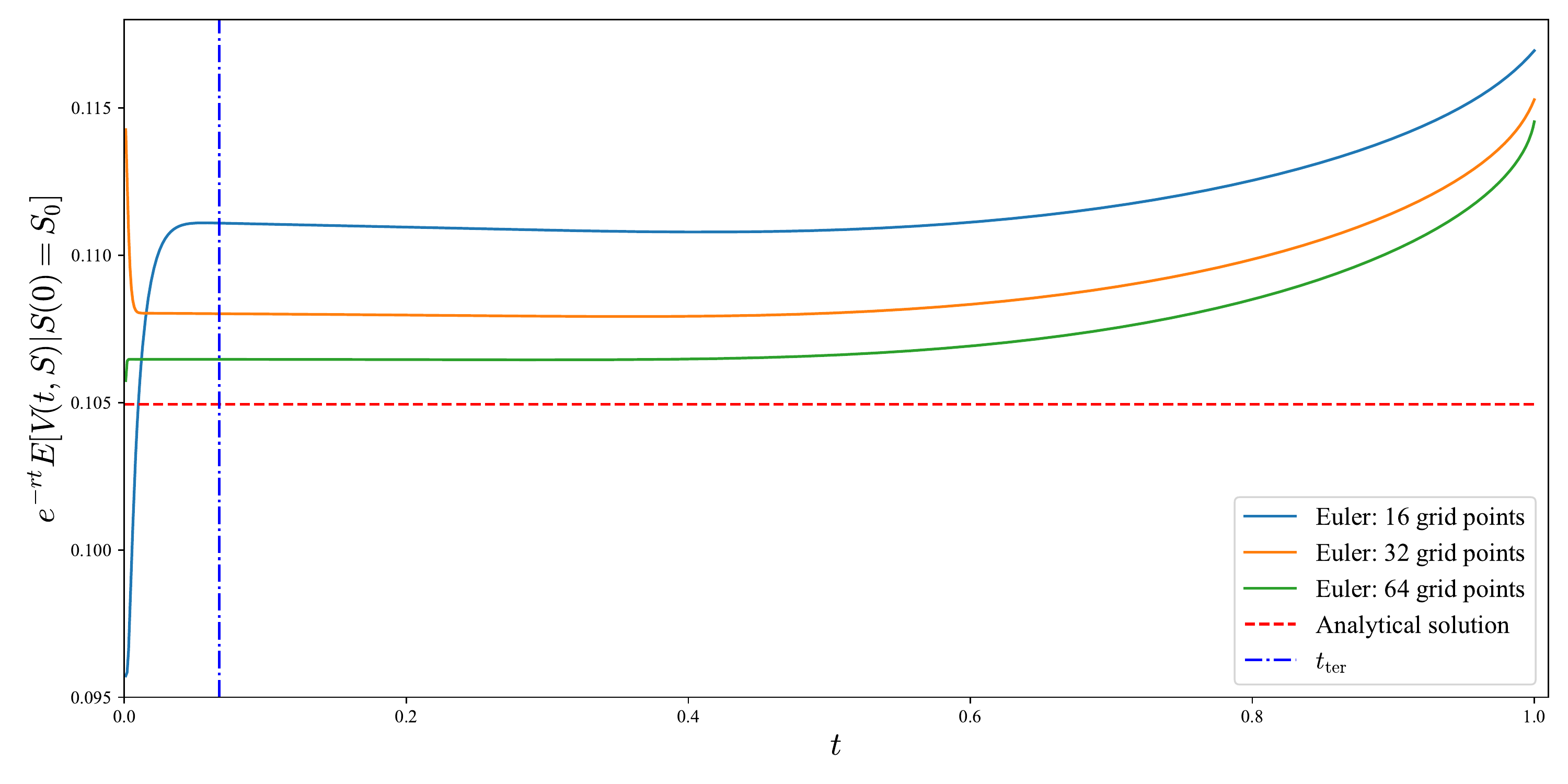}
    \caption{The estimated price of the single-asset double barrier option by classical FDM.}
    \label{fig: classical FDM}
\end{figure}

Before discussing our results, we show the results using the classical FDM in Fig.~\ref{fig: classical FDM}.
The plotted curves are $V(0,S_0) \simeq e^{-rt}E\left[V(t, S)\middle|S(0)=S_0\right]$ at each $t\in[0,T]$, where $V(t, S)$ is calculated by classical FDM and the expectation is taken with respect to the analytical $p(t,s)$.
The error from the analytical solution increases as $t$ increases for $t\geq t_{\mathrm{ter}}$.
This is because, in the range greater than $t_{\mathrm{ter}}$, the probability that the underlying asset price exceeds or falls under the boundary conditions is higher.
As the number of the grid points increases, the derivative price by FDM gets closer to the analytical solution at $t_\mathrm{ter}$.
Since $S_0=1$ is not on the grid points, the error increases when the probability distribution approaches the indicator function with $t\rightarrow0$.

\begin{figure}
    \centering
    \includegraphics[width=\linewidth]{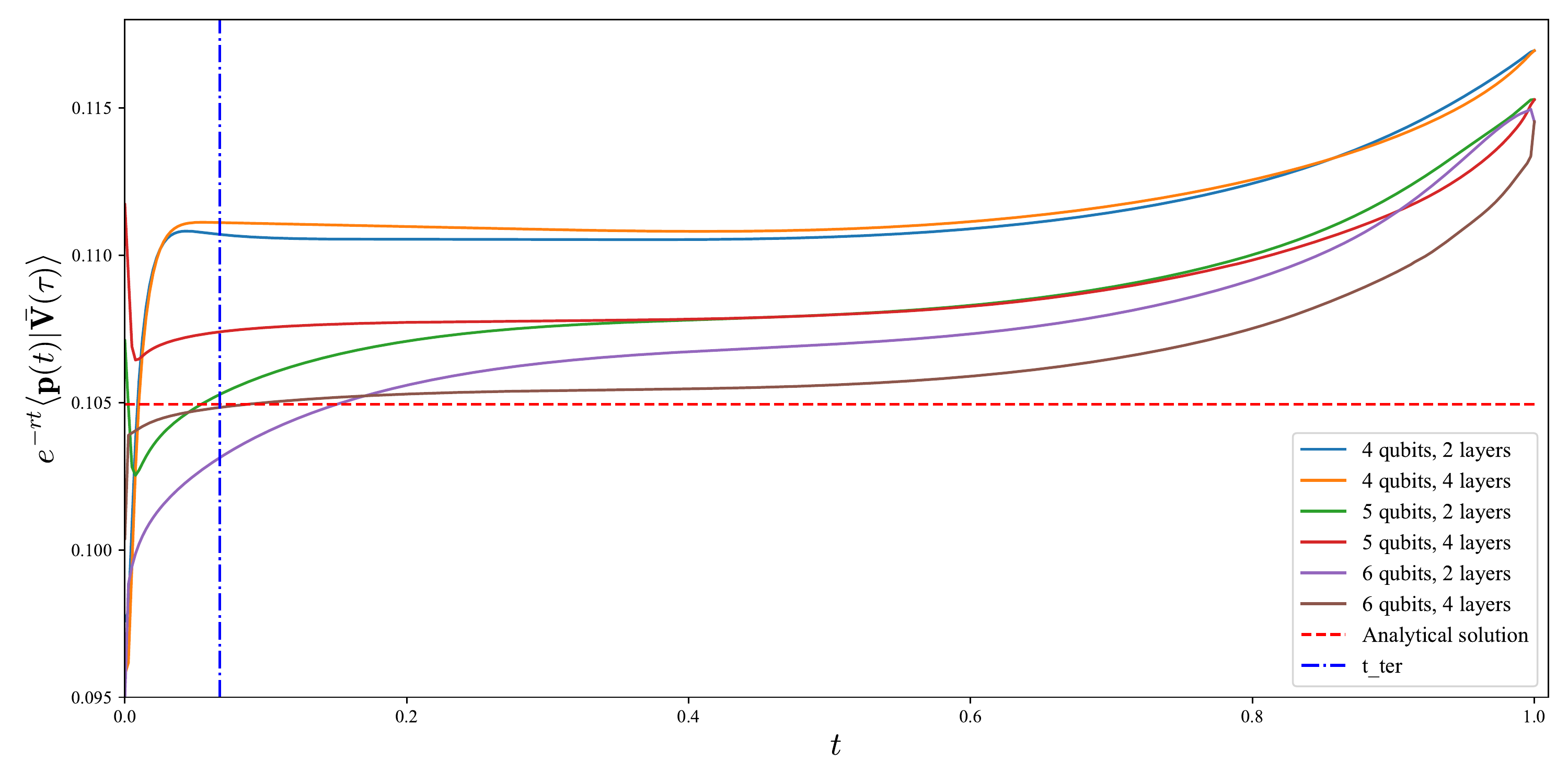}
    \caption{The estimated price of the single-asset double barrier option by the proposed method.}
    \label{fig: VQS}
\end{figure}

Fig.~\ref{fig: VQS} shows the present price of the derivative calculated by our proposed method.
We perform VQS on the simulator and obtain $\ket{\tilde{\bm{v}}(\bm{\theta}(\tau))}$, which is an approximation of $\ket{\bm{\bar{V}}(\tau)}$.
Taking the inner product between $\ket{\tilde{\bm{v}}(\bm{\theta}(\tau))}$ and $\ket{\bm{p}(t)}$, which is calculated by Eqs. (\ref{eq: approximate p}) and (\ref{eq: lognormal pdf}), we obtain the estimation of the present price of the derivative.
In the $4$ qubits case, the result of VQS is a good approximation to the classical FDM solutions of $16$ grid points.
The use of the larger number of qubits, i.e., the larger $n_{\mathrm{gr}}$, gives us solutions that are closer to the analytical solution as in the case of the classical FDM.
In the case of $6$ qubits with $4$ layers, the number of parameters is $30$, which is smaller than the number of grid points of $64$, but the solution is somewhat close to the classical FDM.
Due to computational time requirements, we do not run simulations of larger sizes.
However, we find that the solution obtained with more layers better approximates the classical FDM solution.

\subsection{Possibility of initial state generation}\label{subsec: possibility of initial state generation}
\begin{figure}
    \centering
    \includegraphics[width=\linewidth]{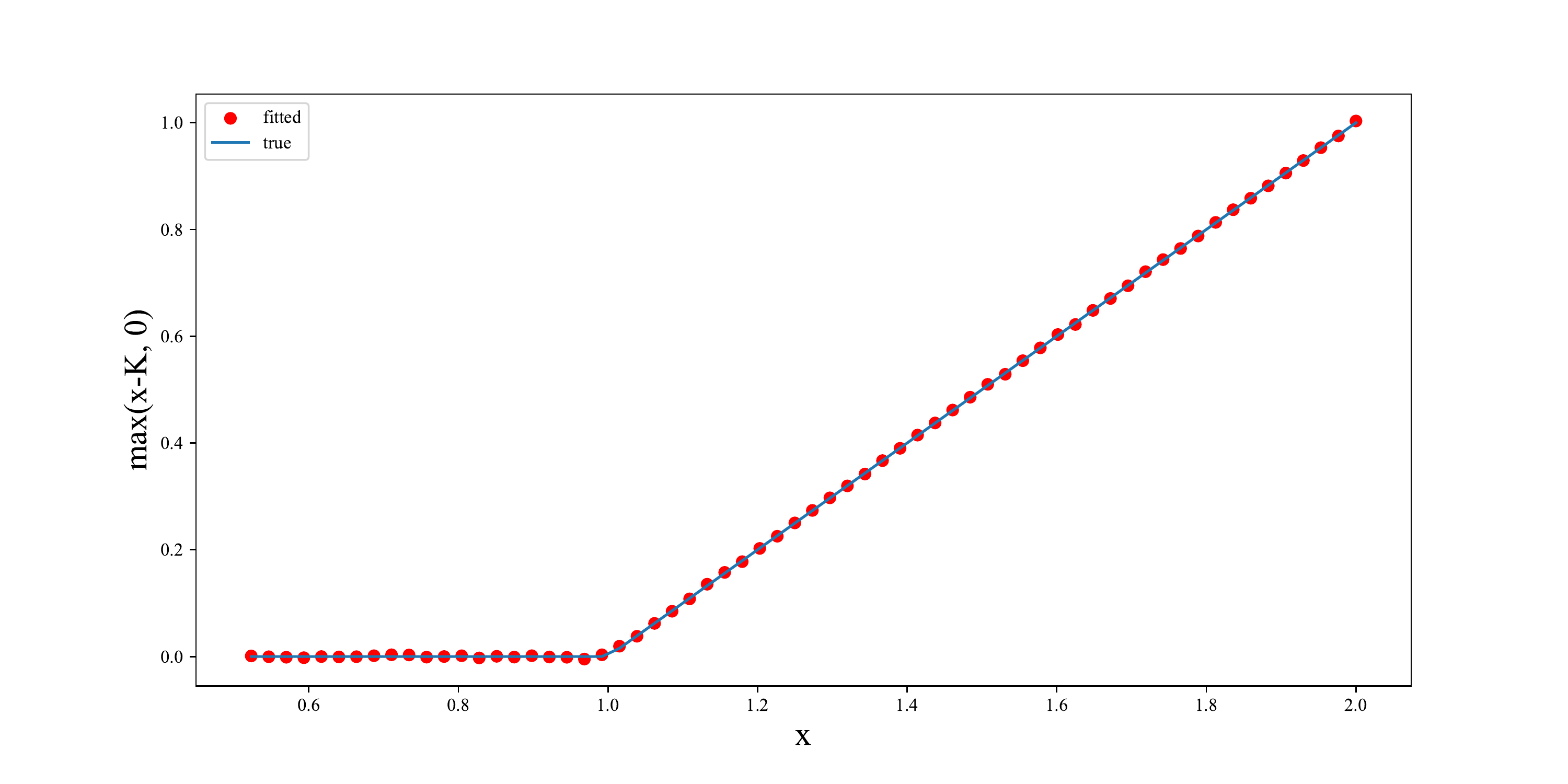}
    \caption{Target initial condition for single assets (solid line) and the initial state obtained by fidelity maximization (circle dots). Parameters are $f_{\mathrm{pay}}(x)=\max\left(x-K, 0\right), K=1$. 
    The value corresponding to fidelity satisfies
    $|1-\alpha_0^{-2}|\braket{\bar{\bm{V}}(0)}{u(\bm{\theta}_0)}|^2|\leq1.2\times10^{-5}$.}
    \label{fig: initial_state}
\end{figure}

To solve the terminal value problem of the BSPDE, it is necessary to prepare the (unnormalized) initial state $\ket{\bar{\bm{V}}(0)} = \sum_kf_{\mathrm{pay}}(\bm{x}^{(k)})\ket{k}$, which we assumed to be given in the previous subsection.
Here, we show by simulation that for a typical $f_{\mathrm{pay}}$, we can approximate the initial state $\ket{\bar{\bm{V}}(0)}$ using an appropriate ansatz.
To show that the initial state can be approximated by $\ket{\nu(\bm{\theta}_0)}=\alpha_0 R_0(\bm{\theta}_0)\ket{0}$, where $\alpha_0=\sqrt{\sum_kf_{\mathrm{pay}}(\bm{x}^{(k)})^2}$ and $R_0(\bm{\theta}_0)$ is the ansatz shown in Fig.~\ref{fig: ansatz}, we adopt L-BFGS-B to find $\bm{\theta}_0$ such that
\begin{align}\label{eq: maximization}
    \max_{\bm{\theta}_0}|\braket{\bar{\bm{V}}(0)}{\nu(\bm{\theta}_0)}|^2,
\end{align}
with SciPy~\cite{2020SciPy-NMeth}.
For the calculation of the gradient, we use the parameter shift rule~\cite{mitarai2018quantum}.
We choose the parameters as $K=1, l=0.5, u=2$ and the ansatz with $6$ qubits and $6$ layers.
By doing maximization of Eq.~\eqref{eq: maximization}, the value $\alpha^{-2}|\braket{\bar{\bm{V}}(0)}{\nu(\bm{\theta}_0)}|^2$, which corresponds to fidelity, should asymptotically converge to $1$.
The result for a payoff function of the single asset call option $f_{\mathrm{pay}}(x)=\max(x-K,0)$ is shown in Fig.~\ref{fig: initial_state}.
We can see that the ansatz approximates the payoff function well.
Indeed, the result satisfies $|1-\alpha_0^{-2}|\braket{\bar{\bm{V}}(0)}{\nu(\bm{\theta}_0)}|^2|\leq 1.2\times10^{-5}$.

Note that this optimization does not correspond to real physical operations.
What we show is that there exists $\bm{\theta}_0$ that at least approximates $\ket{\bar{\bm{V}}(0)}$ well, and we leave the efficient search algorithm for such $\bm{\theta}_0$ to future work.

\section{Conclusion}\label{sec: conclusion}
In this paper, we simulate the BSPDE by VQS and obtain the state which embeds the solution of the BSPDE $\ket{\bm{V}(t_{\mathrm{ter}})}$ at $t_{\mathrm{ter}}$, and utilizing the fact that the derivative price is a martingale, we calculate the derivative price by the inner product of the state $\ket{\bm{V}(t_{\mathrm{ter}})}$ and the state $\ket{\bm{p}(t_{\mathrm{ter}})}$ which embeds the probability distribution.
Although it is difficult to accurately estimate the complexity due to the heuristic nature of variational quantum computation, 
at least in the numerical simulation, we confirm that the proposed method can be performed for the one-asset double barrier option and that the derivative price can be obtained with better accuracy by increasing the number of qubits and the number of layers of ansatz.
We see that the computational complexity is obtained by Table~\ref{tab: complexity} under certain assumptions, and the complexity with respect to $\epsilon$ is $O(1/\epsilon^2\left(\log(1/\epsilon)\right)^d)$.
This means that there would be a significant improvement compared to the classical FDM and conventional quantum algorithms whose complexity has factors like $(1/\epsilon)^{O(d)}$.
Furthermore, we show that an oracle that generates an initial state with embedded payoff functions for typical payoff functions could be represented using an appropriate ansatz.

In this paper, we simply assumed that the initial state of the BSPDE and the state with embedded probability distribution are effectively generated by some variational quantum algorithms. We will confirm this point in future work.

\appendix

\section{Elements of the matrix and the vector of the finite difference method for the BSPDE}\label{app: matrix elements}
Here, we show the concrete elements of $D^{\mathrm{1st}}_{x_i}$ in Eqs.~\eqref{eq:F1st}\eqref{eq:F2nd}, $D^{\mathrm{2nd}}_{x_i}$ in Eq.~\eqref{eq:F2nd}, and $\bm{C}$ in Eq.~\eqref{eq: discretized de}.
$D^{\mathrm{1st}}_{x_i}$ and $D^{\mathrm{2nd}}_{x_i}$ are written by
\begin{align}
    D^{\mathrm{1st}}_{x_i} &= \begin{pmatrix}
        0 & x_i^{(1)} & & & & \\
        -x_i^{(0)} & 0 & x_i^{(2)} & & & \\
        & -x_i^{(1)} & 0 & x_i^{(3)} & & \\
        & & \ddots & \ddots & \ddots & \\
        & & & -x_i^{(n_\mathrm{gr}-3)} & 0 & x_i^{(n_\mathrm{gr}-1)}\\
        & & & & -x_i^{(n_{\mathrm{gr}}-2)} & 0\\
    \end{pmatrix},
\end{align}
and
\begin{widetext}
\begin{align}
    D^{\mathrm{2nd}}_{x_i} &= \begin{pmatrix}
        -2 \left(x_i^{(0)}\right)^2 & \left(x_i^{(1)}\right)^2 & & & & \\
        \left(x_i^{(0)}\right)^2 & -2\left(x_i^{(1)}\right)^2 & \left(x_i^{(2)}\right)^2 & & & \\
        & \left(x_i^{(1)}\right)^2 & -2\left(x_i^{(2)}\right)^2 & \left(x_i^{(3)}\right)^2 & & \\
        & & \ddots & \ddots & \ddots & \\
        & & & \left(x_i^{(n_\mathrm{gr}-3)}\right)^2 & -2\left(x_i^{(n_\mathrm{gr}-2)}\right)^2 & \left(x_i^{(n_\mathrm{gr}-1)}\right)^2\\
        & & & & \left(x_i^{(n_{\mathrm{gr}}-2)}\right)^2 & -2\left(x_i^{(n_\mathrm{gr}-1)}\right)^2\\
    \end{pmatrix},
\end{align}
\end{widetext}
respectively.
$\bm{C}(\tau)$ corresponds to the boundary conditions, and its elements $C_{k}(\tau)$ are
\begin{widetext}
\begin{align}\label{eq: C vector}
    &C_{k}(\tau) = \sum_{i=1}^d \frac{\sigma_i^2}{2h_i^2}\left[(l_i+h_i)^2\delta_{k_i,0}\bar{V}_i^{\mathrm{LB}}(\tau, \bm{x}_{\wedge i}^{(k)}) + (l_i + n_{\mathrm{gr}}h_i)^2\delta_{k_i,n_{\mathrm{gr}}-1} \bar{V}_i^{\mathrm{UB}}(\tau, \bm{x}_{\wedge i}^{(k)})\right] \nonumber\\
    &+ \sum_{i=1}^{d-1}\sum_{j=i+1}^{d}\frac{\sigma_i \sigma_j \rho_{ij}}{4h_ih_j}\nonumber\\
    &\times\left[ -(l_i + h_i)(l_j + (k_j+1)h_j)\delta_{k_i,0}\bar{V}_i^{\mathrm{LB}}(\tau,\bm{x}_{\wedge i}^{(k)}) \right.\nonumber\\
    &\left. - (l_i + (k_i+1)h_i)(l_j + h_j)\delta_{k_j,0}\bar{V}_j^{\mathrm{LB}}(\tau,\bm{x}_{\wedge j}^{(k)})\right.\nonumber\\
    &+\left. (l_i + n_{\mathrm{gr}}h_i)(l_j + (k_j+1)h_j)\delta_{k_i, n_{\mathrm{gr}}-1}\bar{V}_i^{\mathrm{UB}}(\tau,\bm{x}_{\wedge i}^{(k)})\right.\nonumber\\
    &\left. + (l_i + (k_i+1)h_i)(l_j + n_{\mathrm{gr}}h_j)\delta_{k_j,n_{\mathrm{gr}}-1}\bar{V}_j^{\mathrm{UB}}(\tau,\bm{x}_{\wedge j}^{(k)})\right] \nonumber\\
    &+r\sum_{i=1}^d\frac{1}{2h_i}\left[(l_i + n_{\mathrm{gr}}h_i)\delta_{k_i,n_{\mathrm{gr}}-1}\bar{V}_i^{\mathrm{UB}}(\tau,\bm{x}_{\wedge i}^{(k)}) - (l_i + h_i)\delta_{k_i,0} \bar{V}_i^{\mathrm{LB}}(\tau,\bm{x}_{\wedge i}^{(k)})\right],
\end{align}
\end{widetext}
where $\bar{V}_i^{\mathrm{UB}}(\tau,\bm{x}^{(k)}_{\wedge i})=V_i^{\mathrm{UB}}(\tau,\bm{x}^{(k)}_{\wedge i})$ and $\bar{V}_i^{\mathrm{LB}}(\tau,\bm{x}^{(k)}_{\wedge i})=V_i^{\mathrm{LB}}(\tau,\bm{x}^{(k)}_{\wedge i})$.

\section{Decomposition of matrices}\label{app: decomposition}
As discussed in Sec.~\ref{sec:method}, we need to express $F$ and $\ket{\bm{C}(\tau)}$ in terms of linear combination of quantum gates to perform the VQS for the BSPDE.
Here, we show that such decomposition is possible.
The decomposition of $F$ is based on the way shown in Refs.~\cite{kubo2021variational, alghassi2022variational}.
We also obtain a linear combination of quantum gates that generates $\ket{\bm{C}(\tau)}$ by slightly modifying the decomposition of $F$.
For simplicity, we assume $N_{\mathrm{gr}}=2^n$ where $n$ is the number of qubits.
$D^{\mathrm{1st}}_{x_i}, D^{\mathrm{2nd}}_{x_i}$ in Eqs.~\eqref{eq:F1st}\eqref{eq:F2nd} are decomposed as follows,
\begin{align}
    D^{\mathrm{1st}}_{x_i} &= l_i D^{\mathrm{1st}} + h_i \left(\mathrm{Dec}(n)(J(n)+2I^{\otimes n}) \right.\nonumber \\
    & \left.- \mathrm{Inc}(n)\left(J(n) + I^{\otimes n}\right)\right), \\
    D^{\mathrm{2nd}}_{x_i} &= l_i^2 D^{\mathrm{2nd}} \nonumber \\
    &+ 2 l_i h_i \left(\mathrm{Dec}(n)-2 I^{\otimes n} + \mathrm{Inc}(n) \right) (J(n)+I)\nonumber\\
    &+ h_i^2 \left( \mathrm{Dec}(n) -2 I^{\otimes n} + \mathrm{Inc}(n) \right)(J(n)+I)^2.
\end{align}
Here, we define
\begin{align}
    D^{\mathrm{1st}} &\coloneqq -\mathrm{Inc}(n) + \mathrm{Dec}(n) \\
    D^{\mathrm{2nd}} &\coloneqq \mathrm{Inc}(n) + \mathrm{Dec}(n) - 2I^{\otimes n}\\
    J(n) &\coloneqq \sum_{i=0}^{2^n-1} i \ket{i} \bra{i} = \frac{2^n-1}{2} I^{\otimes n} - \sum_{i=1}^n 2^{n-i-1}Z_i \\
    \mathrm{Inc}(n) &\coloneqq \sum_{i=0}^{2^n-2}\ket{i+1}\bra{i}\\
    \mathrm{Dec}(n) &\coloneqq \sum_{i=1}^{2^n-1}\ket{i-1}\bra{i}\\
\end{align}
where $Z_i\coloneqq I^{\otimes i-1}\otimes Z \otimes I^{\otimes n-i}$.
$\mathrm{Inc}(n), \mathrm{Dec}(n)$ are constructed by following operators
\begin{align}
    \mathrm{CycInc}(n) &\coloneqq \sum_{i=0}^{2^n-1} \ket{i+1} \bra{i}, \\
    \mathrm{CycDec}(n) &\coloneqq \sum_{i=1}^{2^n-1} \ket{i-1} \bra{i}, \\
\end{align}
where we define $\ket{-1} \coloneqq \ket{2^n-1}, \ket{2^n}\coloneqq\ket{0}$.
$\mathrm{CycInc}(n), \mathrm{CycDec}(n)$ can be decomposed into a product of $O(n)$ Toffoli, CNOT, X gates with $O(n)$ ancilla qubits~\cite{xiaoyu2014class}.
With these circuits, we obtain
\begin{align}
    \mathrm{Inc}(n)=\frac{1}{2}\mathrm{CycInc}(n)(C^{n-1}Z+I^{\otimes n}),\\
    \mathrm{Dec}(n)=\frac{1}{2}(C^{n-1}Z+I^{\otimes n})\mathrm{CycDec}(n).
\end{align}
$C^{n-1}Z\coloneqq\sum_{i=0}^{2^n-2}\ket{i}\bra{i}-\ket{2^n-1}\bra{2^n-1}$ is an $n$ qubit control Z gate and can be implemented as a product of $O(n^2)$ Toffoli, CNOT, and single-qubit gates~\cite{nielsen2010quantum}.
We can express $D_{x_i}^{\mathrm{1st}}$ and $D_{x_i}^{\mathrm{2nd}}$ as sums of $O(n^2)$ unitary operators, each of which is a product of $O(n^2)$ few-qubit gates.
Then, the first term of Eq.~\eqref{eq:F2nd} is a sum of $O(dn^2)$ operators, each of which is made by $O(n^2)$ few-qubit gates.
The second term is the sum of $O(d^2n^4)$ unitary operators each of which is made by $O(n^2)$ few-qubit gates.
From Eqs.~\eqref{eq:F2nd} and \eqref{eq:F1st}, we see that $F$ can eventually be expressed as a sum of $O(d^2n^4)$ unitary operators each of which is made by $O(n^2)$ few-qubit gates.

It is also necessary to construct a linear combination of unitary operators that outputs the quantum state $\ket{\bm{C}(\tau)}=\sum_{k=1}^{N_{\mathrm{gr}}}C_k(\tau)\ket{k}$.
Here, we consider specific cases where $f_{\mathrm{pay}}(\bm{S}(T))=\max(a_0 + \sum_{i=1}^{d}a_j S_j(T)-K, 0)$, and some assets have knock-out conditions.
These are the cases where the typical boundary conditions introduced in Sec.~\ref{subsec:derivative pricing} are compounded.
In these cases, we can write
\begin{align}
    \ket{\bm{C}(\tau)} = \tilde{G}\ket{0} = 2^{nd/2}G(\tau)H^{\otimes nd}\ket{0}
\end{align}
where
\begin{widetext}
\begin{align}\label{eq: G}
    G(\tau) &= \sum_{i=1}^{d}\frac{\sigma_i^2}{2h_i} \left[(l_i+h_i)^2G^{(0)}_iB^{\mathrm{LB}}_i(\tau)\delta_i^{\mathrm{UB}} + (l_i+n_{\mathrm{gr}}h_i)^2G^{(n_{\mathrm{gr}}-1)}_iB^{\mathrm{UB}}_i(\tau)\delta_i^{\mathrm{UB}}\right]\nonumber\\
    &+ \sum_{i=1}^{d-1}\sum_{j=i+1}^{d}\frac{\sigma_i \sigma_j \rho_{ij}}{4h_ih_j}\nonumber\\
    &\times\left[ -(l_i + h_i)(l_j I^{\otimes dn} + h_jJ_j)G^{(0)}_iB^{\mathrm{LB}}_i(\tau)\delta_i^{\mathrm{LB}}\right.\nonumber\\
    &\left. - (l_i I^{\otimes dn} + h_iJ_i(n))(l_j + h_j)G^{(0)}_jB^{\mathrm{LB}}_j(\tau)\delta_i^{\mathrm{LB}}\right.\nonumber\\
    &+\left. (l_i + n_{\mathrm{gr}}h_i)(l_jI^{\otimes dn} + h_jJ_j)G^{(n_{\mathrm{gr}}-1)}_iB^{\mathrm{UB}}_i(\tau)\delta_i^{\mathrm{UB}}\right.\nonumber\\
    &\left. + (l_i I^{\otimes dn} + h_iJ_i(n))(l_j + n_{\mathrm{gr}}h_j)G^{(n_{\mathrm{gr}}-1)}_jB^{\mathrm{UB}}_j(\tau)\delta_i^{\mathrm{UB}}\right] \nonumber\\
    &+r\sum_{i=1}^d\frac{1}{2h_i}\left[(l_i + n_{\mathrm{gr}}h_i)G^{(n_{\mathrm{gr}}-1)}_iB^{\mathrm{UB}}_i(\tau)\delta_i^{\mathrm{LB}} - (l_i + h_i)G^{(0)}_iB^{\mathrm{LB}}_i(\tau)\delta_i^{LB}\right],
\end{align}
\end{widetext}
where 
\begin{align}
\delta_i^{\mathrm{UB}}&=
    \begin{cases}
    0 & \text{\textit{up and out barrier} is set the $i$-th asset}\\
    1 & \text{otherwise} \\
    \end{cases},
\end{align}
\begin{align}
\delta_i^{\mathrm{LB}}&=
    \begin{cases}
    0 & \text{\textit{down and out barrier} is set to the $i$-th asset} \\
    1 & \text{otherwise} \\
    \end{cases},
\end{align}
and
\begin{align}
    G^{(0)}_i &= I^{\otimes n(i-1)} \otimes \ket{0}\!\bra{0}^{\otimes n}\otimes I^{n(d-i)} \\
    G^{(n_{\mathrm{gr}}-1)}_i &= I^{\otimes n(i-1)} \otimes \ket{1}\!\bra{1}^{\otimes n }\otimes I^{n(d-i)} \\
    B^{\mathrm{UB}}_i(\tau) &= e^{-r\tau}a_0 I^{\otimes nd} \nonumber \\
    &+ \sum_{1\leq j \leq d, j\neq i} a_j \left(l_j I^{\otimes nd} + (n_{\mathrm{gr}}-1) h_j J_j + I^{\otimes nd}\right) \nonumber \\
    &+ a_i l_i I^{\otimes nd}\\
    B^{\mathrm{LB}}_i(\tau) &= e^{-r\tau}a_0 I^{\otimes nd} \nonumber\\
    &+ \sum_{1\leq j \leq d, j\neq i} a_j \left(l_j I^{\otimes nd} + (n_{\mathrm{gr}}-1) h_j J_j + I^{\otimes nd} \right)\nonumber\\
    &+ a_i u_i I^{\otimes nd}\\
    J_i(n)&=I^{\otimes n(i-1)} \otimes (J(n) + I^{\otimes n}) \otimes I^{\otimes n(d-i)}
\end{align}
where $\ket{0}\!\bra{0}^{\otimes n} = \frac{1}{2}\left(I^{\otimes n} - X^{\otimes n} \cdot C^nZ \cdot X^{\otimes n}\right)$ and $\ket{1}\!\bra{1}^{\otimes n} =\frac{1}{2} \left(I^{\otimes n} + C^{n-1}Z\right)$.
$G_i^{(0)}$ and $G_i^{(n_{\mathrm{gr}})}$ are expressed as a sum of $O(1)$ unitary operator each of which is made by $O(n^2)$ few-qubit gates.
$B_i^{\mathrm{UB}}$ and $B_i^{\mathrm{LB}}$ are expressed as a sum of $O(dn)$ unitary operators, each of which is made by $O(n)$ few-qubit gates.
Thus, $G(\tau)$ is a sum of $O(d^2\times n \times dn)=O(d^3n^2)$ unitary operators each of which is made by $O(n^2)$ few-qubit gates.

\section{Variational principle for VQS}\label{app:VQS-additional-term}
Here, we derive Eq.~\eqref{eq:Euler-Lagrange} from a variational principle.
The square of the difference between both sides of Eq.~\eqref{eq:inhomogeneous-ODE} is
\begin{align}\label{eq: norm}
    &\left\| \frac{d}{dt}\ket{\tilde{v}(\bm{\theta}(t))} - L(t)\ket{\tilde{v}(\bm{\theta}(t))} - \ket{\bm{u}(t)}\right\|^2\nonumber\\
    &=\left\|\frac{d}{dt}\ket{\tilde{v}(\bm{\theta}(t))}- L(t)\ket{\tilde{v}(\bm{\theta}(t))} \right\|^2\nonumber \\
    &- 2\Re\left[\bra{\bm{u}(t)}\left(\frac{d}{dt}\ket{\tilde{v}(\bm{\theta}(t))}- L(t)\ket{\tilde{v}(\bm{\theta}(t))} \right)\right] +\left\|^2 \ket{\bm{u}(t)}\right\|^2\nonumber\\
    & =\left\|\frac{d}{dt}\ket{\tilde{v}(\bm{\theta}(t))}\right\|^2 - 2\Re\left[\bra{\tilde{v}(\bm{\theta}(t))}L(t)\frac{d}{dt}\ket{\tilde{v}(\bm{\theta}(t))}\right] \nonumber \\
    &+ \left\|L(t)\ket{\tilde{v}(\bm{\theta}(t))}\right\|^2 \nonumber\\
    & - 2\Re\left[\bra{\bm{u}(t)}\left(\frac{d}{dt}\ket{\tilde{v}(\bm{\theta}(t))}- L(t)\ket{\tilde{v}(\bm{\theta}(t))} \right)\right] +\left\| \ket{\bm{u}(t)}\right\|^2\nonumber\\
    & = 2\Re\sum_{j,k}\frac{d\bra{\tilde{v}(\theta_j(t))}}{d\theta_j(t)}\frac{d\ket{\tilde{v}(\theta_k(t))}}{d\theta_k(t)}\dot{\theta}_j\dot{\theta}_k\nonumber\\
    & - 2\Re\left[\frac{d\bra{\tilde{v}(\bm{\theta}(t))}}{d\theta_j(t)}L(t)\ket{\tilde{v}(\bm{\theta}(t))}+\frac{d\bra{\tilde{v}(\bm{\theta}(t))}}{d\theta_j(t)}\ket{\bm{u}(t)}\right]\dot{\theta}_j  \nonumber\\
    & + 2\Re\left[\bra{\bm{u}(t)}L(t)\ket{\tilde{v}(\bm{\theta}(t))}\right] + \left\|L(t)\ket{\tilde{v}(\bm{\theta}(t))}\right\|^2 +  \left\| \ket{\bm{u}(t)}\right\|^2.
\end{align}
Then, the first order variation of the r.h.s. of Eq.~\ref{eq: norm} is
\begin{align}
    &\delta \left\| \frac{d}{dt}\ket{\tilde{v}(\bm{\theta}(t))} - L(t)\ket{\tilde{v}(\bm{\theta}(t))} - \ket{\bm{u}(t)}\right\|^2\nonumber\\
    &=  2\Re\sum_{j,k}\frac{d\bra{\tilde{v}(\theta_j(t))}}{d\theta_j(t)}\frac{d\ket{\tilde{v}(\theta_k(t))}}{d\theta_k(t)}\dot{\theta}_j\delta\dot{\theta}_k\nonumber\\
    & - 2\Re\left[\frac{d\bra{\tilde{v}(\bm{\theta}(t))}}{d\theta_j(t)}L(t)\ket{\tilde{v}(\bm{\theta}(t))}+\frac{d\bra{\tilde{v}(\bm{\theta}(t))}}{d\theta_j(t)}\ket{\bm{u}(t)}\right]\delta\dot{\theta}_j.
\end{align}
Thus, we obtain Eq.~\eqref{eq:Euler-Lagrange}.

\section{Quantum circuits to evaluate $\mathcal{M}_{i,j}$ and $\mathcal{V}_i$}
\label{app: evaluation of M, V}

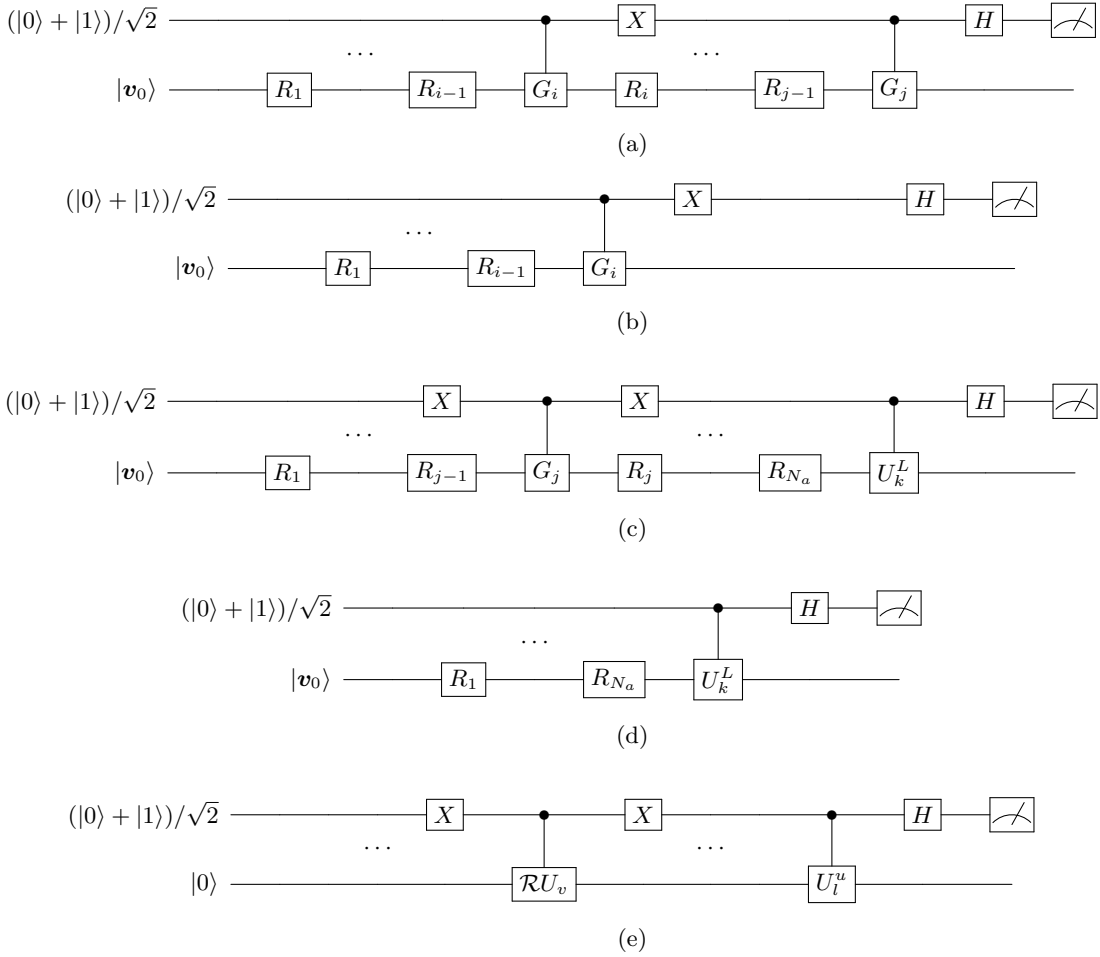
\begin{figure*}[tb]
    \centering
    \begin{equation*}
    \Qcircuit @C=2.0em @R=0.7em {
    \lstick{(\ket{0}+\ket{1})/\sqrt{2}} & \qw & \qw & \qw & \qw &  \ctrl{2}  & \gate{X} &  \qw & \qw & \ctrl{2} & \gate{H} &  \meter\\
    &&&\cdots&&&&\cdots&&&&\\
    \lstick{\ket{\bm{v}_0}} &  \qw &\gate{R_{1}} & \qw & \gate{R_{i-1}} & \gate{G_{i}} &  \gate{R_{i}} & \qw& \gate{R_{j-1}} & \gate{G_{j}} &  \qw & \qw \\
    }
    \end{equation*}(a)
    \begin{equation*}
    \Qcircuit @C=2.0em @R=0.7em {
    \lstick{(\ket{0}+\ket{1})/\sqrt{2}} & \qw & \qw & \qw & \qw &  \ctrl{2}  & \gate{X} &  \qw & \qw & \qw & \gate{H} &  \meter\\
    &&&\cdots&&&&&&&&\\
    \lstick{\ket{\bm{v}_0}} &  \qw &\gate{R_{1}} & \qw & \gate{R_{i-1}} & \gate{G_{i}} &  \qw & \qw& \qw & \qw &  \qw & \qw \\
    }
    \end{equation*}(b)
    \\
    \begin{equation*}
    \Qcircuit @C=2.0em @R=0.7em {
    \lstick{(\ket{0}+\ket{1})/\sqrt{2}} & \qw & \qw & \qw & \gate{X} &  \ctrl{2}  & \gate{X} &  \qw & \qw & \ctrl{2} & \gate{H} &  \meter\\
    &&&\cdots&&&&\cdots&&&&\\
    \lstick{\ket{\bm{v}_0}} &  \qw & \gate{R_{1}} & \qw & \gate{R_{j-1}} & \gate{G_{j}} &  \gate{R_{j}} & \qw& \gate{{R}_{N_a}} & \gate{U^L_{k}} &  \qw & \qw \\
    }
    \end{equation*}(c)
    \\
    \begin{equation*}
    \Qcircuit @C=2.0em @R=0.7em {
    \lstick{(\ket{0}+\ket{1})/\sqrt{2}} &  \qw  & \qw &  \qw & \qw & \ctrl{2} & \gate{H} &  \meter\\
    &&&\cdots&&&&\\
    \lstick{\ket{\bm{v}_0}} &  \qw & \gate{R_{1}} & \qw& \gate{{R}_{N_a}} & \gate{U^L_{k}} &  \qw & \qw \\
    }
    \end{equation*}(d)
    \\
    \begin{equation*}
    \Qcircuit @C=2.0em @R=0.7em {
    \lstick{(\ket{0}+\ket{1})/\sqrt{2}} & \qw & \qw & \qw & \gate{X} &  \ctrl{2}  & \gate{X} &  \qw & \qw & \ctrl{2} & \gate{H} &  \meter\\
    &&&\cdots&&&&\cdots&&&&\\
    \lstick{\ket{0}} & \qw & \qw & \qw & \qw &  \gate{\mathcal{R}U_v} &  \qw & \qw& \qw & \gate{U_l^{u}} &  \qw & \qw \\
    }
    \end{equation*}(e)
    \caption{
      Quantum circuits for evaluating (a) Eq.~\eqref{eq: M1}, (b) Eq.~\eqref{eq: M2}, (c) Eq.~\eqref{eq: V1}, (d) Eq.~\eqref{eq: V1}, and (e) Eqs.~\eqref{eq: V3} and \eqref{eq: V4}. $\mathcal{R}$ is $R_1\cdots R_{i-1}G_{i}R_{i} \cdots R_{N_a}$ for Eq.~\eqref{eq: V3} and $R_1\cdots R_{i-1}G_{i}R_{i} \cdots R_{N_a}$ for Eq.~\eqref{eq: V4} respectively~\cite{endo2020variational}.
    }
    \label{fig:circuitMV}
  \end{figure*}

Here, we show the quantum circuits to evaluate $\mathcal{M}_{i,j}$ and $\mathcal{V}_j$.
Without loss of generality, we can set $i\leq j$.
The terms in Eqs.~\eqref{eq:M} and \eqref{eq:V} are written by
\begin{widetext}
\begin{numcases}{\Re\left(\frac{\partial\bra{\tilde{v}(\bm{\theta}(t))}}{\partial\theta_i} \frac{\partial \ket{\tilde{v}(\bm{\theta}(t))}}{\partial \theta_j}\right) = }
        \theta_0(t)^2\Re\left(\bra{\bm{v}_0}R_{1}^\dagger\cdots R_{i-1}^\dagger G_m^\dagger R_{i}^\dagger \cdots R_{j-1}^\dagger G_j R_{j-1} \cdots R_{1} \ket{\bm{v}_0}\right) & \text{$0 < i \leq j \leq N_a$}\label{eq: M1}\\
        \theta_0(t)\Re\left( \bra{\bm{v}_0}R_{1}^\dagger\cdots R_{j-1}^\dagger G_j^\dagger R_{j-1} \cdots R_{1} \ket{\bm{v}_0} \right)& \text{$0=i<j\leq N_a$}\label{eq: M2} \\
        1 & \text{$m=n=0$}
   \end{numcases}
    \begin{numcases}{\Re\left(\frac{\partial\bra{\tilde{v}(\bm{\theta}(t))}}{\partial\theta_i} U_k^L \ket{\tilde{v}(\bm{\theta}(t)}\right)=}
       \theta_0(t) \Re\left(\bra{\bm{v}_0}R_{1}^\dagger\cdots R_{i-1}^\dagger G_i^\dagger R_{i}^\dagger \cdots R_{N_a}^\dagger U_k^L R_{N_a}\cdots R_1 \ket{\bm{v}_0}\right) & \text{ $i\neq 0$}\label{eq: V1}\\
       \Re\left(\bra{\bm{v}_0}R_{1}^\dagger \cdots R_{N_a}^\dagger {U_k^L}^\dagger R_{N_a} \cdots R_{1} \ket{\bm{v}_0}\right) & \text{$i=0$} \label{eq: V2}
   \end{numcases}
   \begin{numcases}{\Re\left(\frac{\partial\bra{\tilde{v}(\bm{\theta}(t))}}{\partial\theta_m} U_l^{u} \ket{0}\right)=}
       \theta_0(t)\Re\left( \bra{\bm{v}_0}R_{1}^\dagger\cdots R_{i-1}^\dagger G_m^\dagger R_{i}^\dagger \cdots R_{N_a}^\dagger U_l^{u} \ket{0}\right) & \text{$i\neq 0$}\label{eq: V3}\\
       \Re\left(\bra{\bm{v}_0}R_{1}^\dagger\cdots R_{N_a}^\dagger {U_l^{u}}^\dagger \ket{0}\right) & \text{ $i=0$} \label{eq: V4}
   \end{numcases}.
\end{widetext}
We can evaluate these terms using quantum circuits depicted in Fig.~\ref{fig:circuitMV}.
Note that, although the quantum circuit evaluating Eqs.~\eqref{eq: V3} and \eqref{eq: V4} contains the control-$\mathcal{R}U_v$ gate, where 
\begin{align}\label{eq: R}
    \mathcal{R}=R_1\cdots R_{m-1}G_{m}R_{m} \cdots R_{N_a}
\end{align}
for Eq.~\eqref{eq: V3} and $R_1\cdots R_{i-1}G_{i}R_{i} \cdots R_{N_a}$ for Eq.~\eqref{eq: V4} respectively, in the case where all boundary conditions are knock-out barriers, that is, in the case of $\ket{\bm{C}(t)}=\bm{0}$, we do not need to evaluate Eqs.~\eqref{eq: V3} and \eqref{eq: V4}.

\section{Lower bound of $\Xi$}\label{app: lower bound of Xi}
Here, we evaluate the lower bound of $\Xi$ and show that the $\Xi$ does not decrease exponentially with respect to the number of assets $d$.
Using the inequality
\begin{align}
    \min(a^2, b^2) \leq \frac{1}{4}(a+b)^2 
\end{align}
for $a,b\in\mathbb{R}_+$, we obtain
\begin{align}\label{eq: upper bound of t_ter}
    t_{\text {ter }} &\leq \min \left\{\frac{2\left(\log \frac{u_{i}}{s_{i, 0}}\right)^{2}}{25 \sigma_{i}^{2} \log \frac{2 \tilde{A} d(d+1)}{\epsilon}}, \frac{2\left(\log \frac{s_{i, 0}}{l_{i}}\right)^{2}}{25 \sigma_{i}^{2} \log \frac{2 \tilde{A} d(d+1)}{\epsilon}}\right\} \nonumber\\
    &\leq \frac{\left(\log \frac{u_{i}}{l_{i}}\right)^{2}}{50 \sigma_{i}^{2} \log \frac{2 \tilde{A} d(d+1)}{\epsilon}},
\end{align}
for $i\in[d]$.
From Eqs.~\eqref{eq: gamma}\eqref{eq: upper bound of t_ter}, $\Xi$ is evaluated by
\begin{align}
    \Xi \geq \zeta B^2\left(\frac{25}{4 \pi} \log \frac{2 \tilde{A} d(d+1)}{\epsilon}\right)^{d / 2} \prod_{i=1}^{d} \frac{\frac{u_{i}}{l_{i}}-1}{\log \frac{u_{i}}{l_{i}}}
\end{align}
As easily verified by elementary analysis, for any $z > 1$,
\begin{align}
    \frac{z-1}{\log z}\geq 1
\end{align}
holds, and then, we obtain
\begin{align}
    \Xi \geq \zeta B^2\left(\frac{25}{4 \pi} \log \frac{2 \tilde{A} d(d+1)}{\epsilon}\right)^{d / 2}.
\end{align}
Thus, we can see that $\Xi$ does not decrease exponentially with respect to $d$.

\begin{acknowledgments}
Koichi Miyamoto is supported by JSPS KAKENHI Grant No. 22K11924.
Kosuke Mitarai is supported by JST PRESTO Grant No. JPMJPR2019 and JSPS KAKENHI Grant No. 20K22330.
Keisuke Fujii is supported by JST ERATO JPMJER1601, and JST CREST JPMJCR1673.
This work is supported by MEXT Quantum Leap Flagship Program (MEXT QLEAP) Grant Number JPMXS0118067394 and JPMXS0120319794.
We also acknowledge support from JST COI-NEXT program.
\end{acknowledgments}
\bibliography{bibliography}
\bibliographystyle{unsrt}
\end{document}